\documentclass[notitlepage,aps,prl,reprint,twocolumn,longbibliography,superscriptaddress]
{revtex4-1}
\usepackage{graphicx}
\usepackage{amsmath}
\usepackage{amssymb}
\usepackage{comment}
\usepackage[colorlinks, allcolors=blue]{hyperref}
\usepackage[all]{hypcap}
\usepackage[mathlines]{lineno}
\usepackage{physics}
\usepackage{wrapfig}
\usepackage{lipsum}
\usepackage[normalem]{ulem}
\usepackage{siunitx}
\usepackage{bbold}
\usepackage{animate}

\newcommand{\pref}[2]{\hyperref[#1]{\ref{#1}(#2)}}
\newcommand{\preff}[2]{\hyperref[#1]{\ref{#1}#2}}
\newcommand{\eqpref}[1]{\hyperref[#1]{(\ref{#1})}}

\newcommand{\squig}{{\raise.17ex\hbox{$\scriptstyle\sim$}}}

\begin{document}
	\title{Interaction-enabled topological pumping of Rydberg electrons}
\author{Chenxi Huang}
	\thanks{These authors contributed equally to this work.}   
\affiliation{Department of Physics, University of Illinois at Urbana-Champaign, Urbana, IL 61801-3080, USA}
\affiliation{Department of Physics, The Pennsylvania State University, University Park, Pennsylvania 16802, USA}  
\author{Tao Chen}
	\thanks{These authors contributed equally to this work.}
\affiliation{School of Physics, Xi’an Jiaotong University, Xi’an 710049, China}
\affiliation{Department of Physics, University of Illinois at Urbana-Champaign, Urbana, IL 61801-3080, USA}
\affiliation{Department of Physics, The Pennsylvania State University, University Park, Pennsylvania 16802, USA}
 \author{Kaden R.~A. Hazzard}
\affiliation{Department of Physics and Astronomy, Rice University, Houston, TX 77005, USA}
        \affiliation{Smalley-Curl Institute, Rice University, Houston, TX 77005, USA}
	\author{Jacob P. Covey}
	\affiliation{Department of Physics, University of Illinois at Urbana-Champaign, Urbana, IL 61801-3080, USA}
	\author{Bryce Gadway}
\email{bgadway@psu.edu}
	\affiliation{Department of Physics, University of Illinois at Urbana-Champaign, Urbana, IL 61801-3080, USA}
 \affiliation{Department of Physics, The Pennsylvania State University, University Park, Pennsylvania 16802, USA}
	\date{\today}

\begin{abstract}
Topological pumping is a paradigmatic realization of quantized transport in band systems, yet its fate in strongly correlated regimes, especially with long-range interactions, remains largely unexplored. Here we report the experimental observation of interaction-enabled topological pumping of correlated Rydberg electrons in a synthetic lattice. We show that dipolar exchange interactions induce a controllable shift of the underlying topological singularity in parameter space, such that a fixed pumping trajectory can be driven through successive topological transitions by tuning the interaction strength alone. This leads to the emergence and breakdown of quantized transport. The observations are consistent with an effective Rice–Mele description with interaction-renormalized onsite potentials and are supported by characterizing the adiabaticity and robustness to control trajectory imperfections. Our results establish a platform for exploring interaction-controlled topological transport beyond perturbative regimes and open a route toward engineering correlated topological matter in synthetic quantum systems.
\end{abstract}

\maketitle

Strongly interacting topological quantum matter exhibits essentially exotic behaviors with potential implications and applications for quantum science and technology~\cite{Haldane2017,Chetan2008,Wen2004,Cooper2019}. Understanding the interplay between nontrivial topology and interactions is crucial for interpretations of many fundamental phenomena, from the perturbative robustness of quantized charge transport~\cite{thoulessQuantizationParticleTransport1983,Niu1984,citroThoulessPumpingTopology2023,lohseThoulessQuantumPump2016,nakajimaTopologicalThoulessPumping2016,Lu2016} to the emergence of fractional quantum Hall states~\cite{Laughlin1983,Hafezi2007,Clark2020,Leonard2023,Bohm2026}. Significant theoretical efforts have therefore been made to identify the underlying topological invariants that govern many-body pumping dynamics for strongly correlated systems~\cite{Kuno2020, kuno2024topologicaldomainwallpumpmathbbz2,yanYangMonopolesEmergent2018,Lam2024,Mostaan2022,Zeng2015,Zeng2016,Berg2011,Taddia2017,Nakagawa2018,Unanyan2020,Chenyl2020,Fu2022a,Fu2022b,Ke2017,Huang2024,Wu2025,Jurgensen2022,Jurgensen2025,Tao2025}. Experimentally, interaction effects on topological pumping have recently attracted increasing attention in various platforms. In nonlinear photonic waveguide systems, both quantized and fractional topological pumping of solitons have been observed~\cite{jurgensenQuantizedNonlinearThouless2021,jurgensenQuantizedFractionalThouless2023,Chaudhari2025}. In parallel, Hubbard interactions have also been shown to either destroy or enable many-body topological pumping for correlated fermions in optical lattices~\cite{walterQuantizationItsBreakdown2023,viebahnInteractionsEnableThouless2024,Zhu2024,Kiefer2026}. 

A key ingredient to understanding the topological properties of correlated particles is that interactions can effectively shift the gapless critical point, i.e., the topological singularity, in the parameter space of the Rice-Mele lattice \cite{yanYangMonopolesEmergent2018,viebahnInteractionsEnableThouless2024}. In the absence of interactions, the singularity is located at the band-crossing point and acts as the source of Berry curvature responsible for quantized transport. When the pumping trajectory encloses the singularity, the transported charge per cycle is topologically quantized \cite{thoulessQuantizationParticleTransport1983,Xiao2010}. Weak interactions perturbatively split or shift the singularity while preserving the overall topological structure, such that the singularity remains inside the modulation loop and the pumping remains quantized \cite{Niu1984,yanYangMonopolesEmergent2018,Xiao2010,Essin2011}. When continuously tuning the interaction strength, the singularity can move into and out of the pumping trajectory, naturally leading to interaction-driven topological phase transitions accompanied by the emergence and breakdown of the charge pump. Such mechanisms have been demonstrated in Hubbard many-body systems with tunable short-range contact interactions \cite{walterQuantizationItsBreakdown2023,viebahnInteractionsEnableThouless2024}. However, whether a similar picture survives in systems with nonlocal long-range interactions, such as the dipolar exchange interaction between Rydberg atoms, remains largely unexplored, particularly in the few-body regime where interaction effects are intrinsically prominent.

Atom arrays provide a clean and versatile platform for studying topological transport of strongly correlated Rydberg electrons in the few-body limit. Here, we show a clear picture of the interaction-enabled topological pumping when the designed parameter trajectory wraps around the interaction-shifted singularity for a pair of Rydberg electrons, in a regime where individual Rydberg electrons are not pumped. Different from our previous work \cite{Huang2025}, here a two-color microwave driving protocol is employed to selectively address collective transitions via the intermediate triplets under the pair basis \cite{Chen2024a}. For a given fixed pumping path, we observe two topological phase transitions when tuning the interaction from the non-interacting limit to the strongly correlated regime. We find that quantized directional transport emerges only within an intermediate interaction regime by monitoring the charge moving after one modulation cycle. We further analyze the adiabatic condition and examine the robustness against perturbations for such interaction-induced pumping dynamics. 

\begin{figure*}[]
	\includegraphics[width=\textwidth]{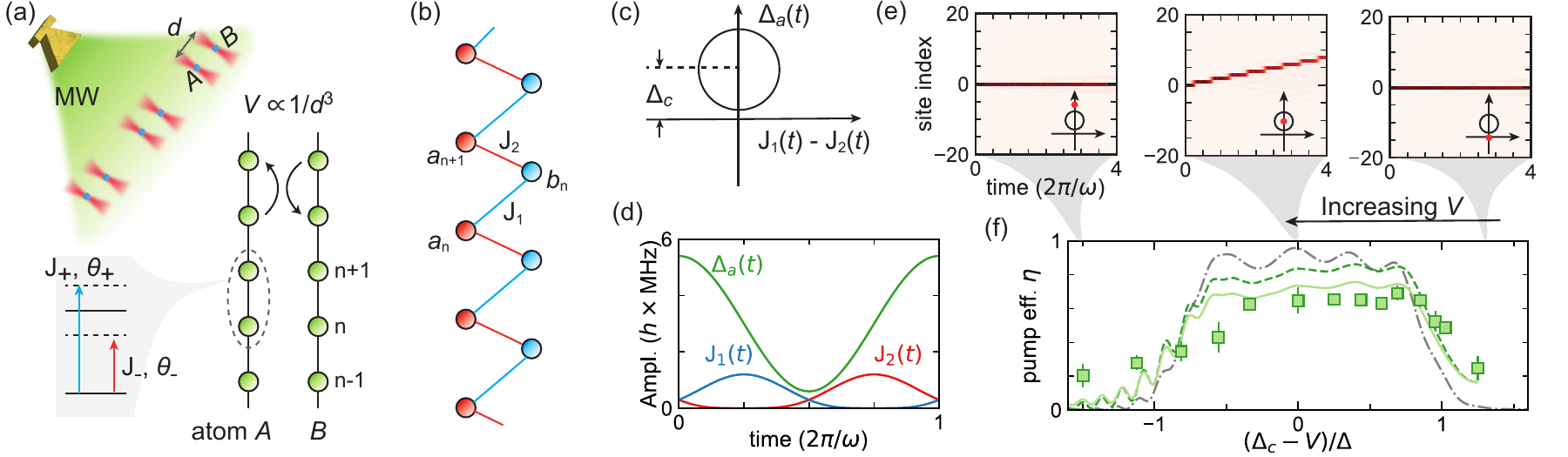}
	\caption{
    \textbf{Experimental scheme for interaction-enabled topological pumping of correlated Rydberg electrons.} 
    \textbf{(a)} An array of independent pairs of two Rydberg atoms prepared with optical tweezers ($A$ and $B$) with tunable separation $d$ are driven by multi-frequency microwave fields, forming a synthetic lattice of Rydberg states $\ket{n}$. Each nearest-neighbor transition is coupled by two microwave tones with time-dependent amplitudes $J_\pm (t)$ and phases $\theta_\pm(t)$. Dipolar exchange interactions scale as $V_n \approx V \propto 1/d^3$. 
    \textbf{(b)} Effective Rice-Mele representation in the pair basis $\ket{a_n}=\ket{n}_A\otimes\ket{n}_B=\ket{n,n}$ and $\ket{b_n}=\ket{+}_n=(\ket{n}_A\otimes\ket{n+1}_B+\ket{n+1}_A\otimes\ket{n}_B)/\sqrt{2}$, with the intra- and inter-cell hopping amplitudes $J_{1,2}=\sqrt{2}J_\pm$. 
    \textbf{(c)} Pumping trajectory in parameter space, with onsite offset $\Delta_a(t)$ for site $\ket{a_n}$ centered at $\Delta_c$ and modulation amplitude $\Delta$. 
    \textbf{(d)} Experimentally used time-dependent driving parameters $J_{1,2}(t)$ and $\Delta_a(t)$ with $J_0/h = 0.3~{\rm MHz}$, $\Delta_c/h=3~{\rm MHz}$ and $\Delta/h=2.4~{\rm MHz}$. Here $\omega$ is the modulation frequency. 
    \textbf{(e)} Numerical dynamics of an initially localized state $\ket{a_0}$ from the effective Hamiltonian (\ref{eq2}) under different interaction strength: from left to right, $V-\Delta_c=\{1.5\Delta, 0, -1.25\Delta\}$. Here we consider an 81-site chain, with even (odd) index for $\ket{a_n}$ ($\ket{b_n}$), $\Delta_c/J_0=100$, $\Delta/J_0=8$, $\hbar\omega/J_0=1$. Red marks indicate the singularity positions. 
    \textbf{(f)} Pumping efficiency $\eta$ vs. the interaction strength $V$ under a fixed pumping trajectory. Experimental data (green squares) are compared with simulations of the effective Rice–Mele model (dotted dashed line, with settings in (e)) and the full Hamiltonian, with and without state-preparation inefficiency (solid and dashed lines, respectively, with experimental parameters in (d)). The error bars are standard errors from multiple independent experimental data sets.}
\label{fig1}
\end{figure*}

Figure \pref{fig1}{a} shows the experimental implementation scheme. We consider a pair of single Rydberg atoms, $A$ and $B$, with an interatomic distance of $d$ tunable by adjusting the optical tweezers. As in previous studies \cite{Huang2025,Chen2024a,Chen2024,Chen2025,chen2025b,Kanungo2022,Trautmann2024}, a set of Rydberg states $\ket{n}$ is coupled by multi-frequency microwave (MW) fields to form a synthetic Rydberg lattice. But here, each nearest-neighbor transition is simultaneously driven by two tones \cite{Chen2024a} with independently controlled amplitudes $J_\pm(t)$ and phases $\theta_\pm(t)$. In the presence of state-dependent dipolar exchange interactions $V_n\propto 1/d^3$, the system is described by \cite{SuppMats}
\begin{eqnarray}\label{eq1}
 H_{\rm int} &=& \sum_n\sum_{\alpha\in\{A, B\}} (J_+ e^{i\theta_+} + J_- e^{i\theta_-})\ket{n}_\alpha\bra{n+1} \nonumber\\
 &+& \sum_n V_n \ket{n+1}_A\bra{n} \otimes \ket{n}_B\bra{n+1} + {\rm H.c.},
\end{eqnarray}
where the phase modulation $\theta_\pm (t)= \pm \Delta_c t \pm \Delta \sin{(\omega t)}/\omega$ introduces both a static frequency shift $\pm \Delta_c$ and a periodically varied detuning with the amplitude $\Delta$ and frequency $\omega$, and $J_\pm (t) = J_0 \left[1 \pm \sin{(\omega t)}\right]^2/2\sqrt{2}$.

To illustrate the role of interactions in the pumping process, we move to the pair basis consisting of the product states $\ket{a_n}=\ket{n}_A\otimes\ket{n}_B$ and the pair triplets $\ket{b_n}=\ket{+}_n=(\ket{n}_A\otimes\ket{n+1}_B+\ket{n+1}_A\otimes\ket{n}_B)/\sqrt{2}$ (Here the asymmetric singlets are fully decoupled); see Fig.~\pref{fig1}{b}.  Given that $\Delta_c$ dominates the energy scale, a local gauge transformation followed by a rotating-wave approximation \cite{SuppMats} maps the system into an effective pair-state Rice-Mele Hamiltonian 
\begin{eqnarray}\label{eq2}
 H_{\rm eff} &=& \sum_n\left[J_1(t)a_n^\dagger b_n + J_2(t) b_n^\dagger a_{n+1}+{\rm H.c.}\right] \nonumber\\
 &+& \sum_n \left[\Delta_a (t) a_n^\dagger a_n + V_n b_n^\dagger b_n\right],
\end{eqnarray}
where $J_{1,2}(t) = \sqrt{2}J_\pm (t)$, and $\Delta_a (t) = \Delta_c + \Delta \cos{(\omega t)}$. In this representation, the dipolar exchange interaction appears as an effective onsite energy shift of the triplet states $\ket{b_n}$, and therefore varying the interaction strength directly shifts the position of the topological singularity in parameter space.

Figure~\pref{fig1}{c} shows the modulation trajectory in the non-interacting limit. With $\Delta_c > \Delta$, the singularity lies outside the modulation path, and consequently the Rydberg electrons initial in the $\ket{a_0}$ site almost freeze, i.e., no charge pumping happens [Fig.~\pref{fig1}{e}, right panel]. As the interaction strength increases, the singularity is displaced toward the trajectory. For $V_n\approx V = \Delta_c$, the singularity lies at the center of the pumping loop, rendering the pumping process topologically nontrivial and inducing directional transport of the correlated Rydberg electron pair [Fig.~\pref{fig1}{e}, middle panel]. Upon further increasing the interaction strength, the singularity exits the modulation loop again, suppressing the transport and restoring a topologically trivial regime [Fig.~\pref{fig1}{e}, left panel]. These case studies suggest that interaction-enabled topological phase transitions occur two times as the interaction strength is tuned. Figure \pref{fig1}{f} summarizes the pumping efficiency indicated by the population in $\ket{1}$ after one modulation period under different interaction strengths. Effective pumping happens only within the interval bound by $V=\Delta_c \pm \Delta$ (where the singularity crosses the trajectory boundary), where the topology is non-trivial and the pumping efficiency is topologically protected by the quantized Chern number. 

Experimentally, we verify this interaction-enabled pumping mechanism in our $^{39}{\rm K}$ tweezer platform, using five Rydberg states, $\{42S_{1/2}, \allowbreak 42P_{3/2}, \allowbreak 43S_{1/2}, \allowbreak 43P_{3/2}, \allowbreak 44S_{1/2}\}$ all with $m_J=1/2$, to encode the synthetic lattice sites from $\ket{n=0}$ to $\ket{4}$. Atom pairs are initialized in the $\ket{0}=\ket{42S_{1/2}}$ state via two-photon Rydberg excitations \cite{Chen2024}. The periodic modulation functions for each parameter are shown in Fig.~\pref{fig1}{d}. The dipolar exchange interactions between neighboring $S$ and $P$ states are nearly uniform, with small variations arising from the corresponding different $C_3$ coefficients \cite{SuppMats}. The overall interactions are globally tuned by varying $d$ and characterized by $V\equiv V_0=-C_3^{01}/(2d^3)$ with $C_3^{01}/h = -1502~{\rm MHz~\mu m^3}$. State populations, $P_n = \langle \hat{P}_n \rangle$ with $\hat{P}_n=(\ket{n}_A\bra{n}\otimes\mathbb{1}_B + \mathbb{1}_A\otimes\ket{n}_B\bra{n})/2$, are measured by de-excitation of the atoms to the ground state followed by fluorescence imaging. Throughout the experiment, we postselect events in which both optical tweezers are successfully loaded. The measured populations are renormalized based on independently determined baselines to account for the two-photon Rydberg excitation inefficiency and the finite lifetime of the Rydberg states~\cite{Chen2024,SuppMats}.

\begin{figure}[b]
	\includegraphics[width=0.48\textwidth]{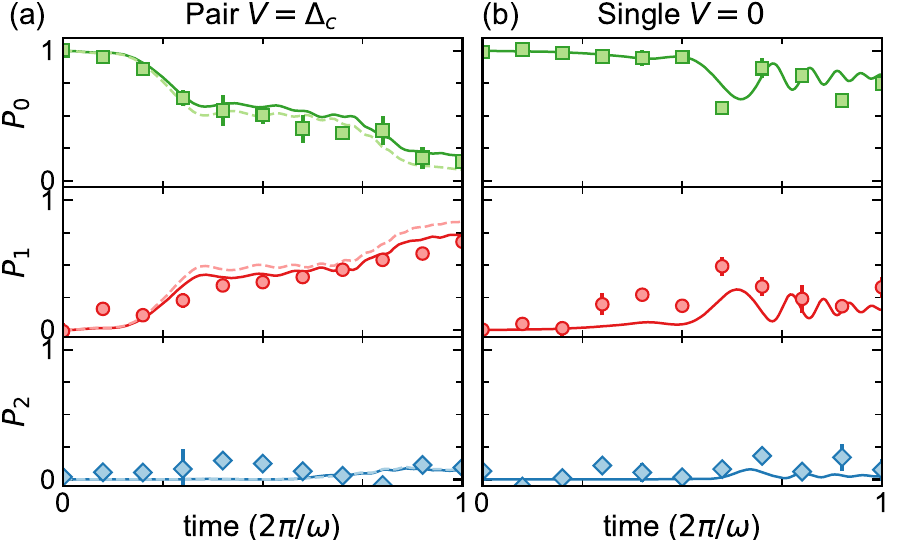}
	\caption{
    \textbf{Pumping dynamics for Rydberg electrons in synthetic lattices.} 
    Measured population $P_n$ in each of the synthetic sites $\ket{n}$ over one modulation period for \textbf{(a)} correlated pairs with $V=\Delta_c$ and \textbf{(b)} uncorrelated singles with $V=0$. The pumping parameters used are: $\Delta_c/h = 3.0~{\rm MHz}$, $\Delta/h = 2.4~{\rm MHz}$, $J_0/h = 0.3~{\rm MHz}$, and $\omega/2\pi = 0.3~{\rm MHz}$. The solid lines in each panel are numerical simulation results with the original Hamiltonian (\ref{eq1}), by taking into account the initial state preparation inefficiency. The dashed lines in (a) indicate simulation results by assuming perfect initial state preparation. The error bars are standard errors from multiple independent experimental data sets. 
    }
\label{fig2}
\end{figure}

We first examine the pumping dynamics of two correlated Rydberg electrons. As illustrated in Fig.~\ref{fig2}, for the chosen trajectory with $\Delta = 0.8\Delta_c$, uncorrelated Rydberg electrons ($V = 0$) exhibit negligible net transport. The population remains localized in the initial state, consistent with a topologically trivial regime where the modulation path does not enclose the singularity [Fig.~\pref{fig1}{e}, right panel]. However, for $V=\Delta_c$, the correlated Rydberg electrons move together from initial $\ket{n=0}$ to $\ket{n=1}$ state after one modulating cycle, corresponding to unit-cell shift from $\ket{a_0}$ to $\ket{a_1}$ under the mapped effective Rice-Mele picture in Fig.~\pref{fig1}{b}. The dynamics reveal a two-step pumping process within a single modulation period. During the first half-cycle, the Rydberg electron pair moves into the triplet $\ket{b_0}$, with $P_0$ and $P_1$ are both approximately 0.5; see Fig.~\pref{fig2}{a}. In the second half-cycle, the electron pair undergoes the transition from $\ket{b_0}$ to $\ket{a_1}$, accompanied by an increase of $P_1$ and a corresponding depletion of $P_0$. 

To quantify the interaction dependence under a fixed modulation path, we use the population in state $\ket{n=1}$ after one full modulation period to characterize the pumping efficiency, i.e., $\eta = P_1(t=2\pi/\omega)$. As shown in Fig.~\pref{fig1}{f}, the measured $P_1$ rapidly grows as the interaction strength increases such that the singularity crosses the lower boundary ($V=\Delta_c-\Delta$) of the pumping trajectory. In the intermediate regime, where the singularity is enclosed by the modulation loop, the efficiency remains nearly constant, reflecting topologically protected transport. For larger interaction strength, the singularity exits the loop at $V=\Delta_c+\Delta$, and the pumping efficiency decreases, signaling a breakdown of topological protection.

The measured interaction dependence is in good agreement with numerical simulations based on both the original full Hamiltonian (\ref{eq1}) and the effective Rice–Mele model (\ref{eq2}). The remaining quantitative deviations arise primarily from two effects. First, the mapping to the effective Rice–Mele Hamiltonian neglects: (i) fast-oscillating terms that become relevant at finite $\Delta_c$, and (ii) the leakage processes to other pair states \cite{SuppMats}. On one hand, the two microwave tones contribute simultaneously to both intra- and inter-cell couplings, leading to residual corrections beyond the rotating-wave approximation. On another, the transition from $\ket{+}_0$ to $(\ket{0}_A\ket{2}_B+\ket{2}_A\ket{0}_B)/\sqrt{2}$ messes up the pumping dynamics, as evidenced by the simulation result with the full Hamiltonian showing a slight increase of $P_2$ to $\sim 0.1$ over one modulation period [Fig.~\pref{fig2}{a}]. These account for the difference between simulations based on the full and effective Hamiltonians shown in in Fig.~\pref{fig1}{f} as well as the degrade of the pumping efficiency from the quantized value. Second, the finite initial-state preparation fidelity leads to n admixture of atoms remaining in the atomic ground state, thus appearing to remain in state $\ket{n=0}$ based on our measurement protocol \cite{Chen2024}. From independent calibration of the optical pumping and Rydberg excitation sequence, approximately 15\% events do not contribute to the correlated pair pumping dynamics~\cite{Chen2024}. Including this imperfection in the numerical simulations yields excellent agreement with the measured data. Other effects, such as the finite-size of our synthetic Rydberg lattice and the nearly but not perfectly uniform state-dependent interactions, would also modify the perfect quantized value but do not change the overall interaction-driven transitions or the underlying topological mechanism.

\begin{figure}[]
	\includegraphics[width=0.48\textwidth]{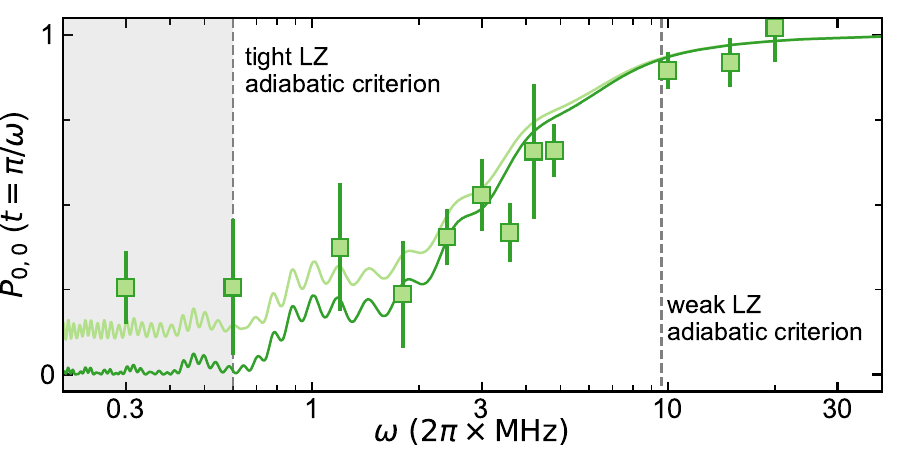}
	\caption{
    \textbf{Adiabaticity of the interaction-enabled topological pumping.} 
    Measured population $P_{0,0}(t=\pi/\omega)$ in the initial $\ket{a_0}=\ket{0,0}$ state as a function of the modulating frequency $\omega$. The pumping parameters used are: $V/h=\Delta_c/h = 3.0~{\rm MHz}$, $\Delta/h = 2.4~{\rm MHz}$, and $J_0/h = 0.3~{\rm MHz}$. The solid light and dark lines are numerical simulation results with and without the consideration of the initial state preparation inefficiency, respectively. The dashed vertical lines indicate the positions of critical values of $\omega$ derived from the Landau-Zener (LZ) adiabatic conditions \cite{SuppMats}. The error bars are standard errors from multiple independent experimental data sets. 
    }
\label{fig3}
\end{figure}

Next we study the adiabatic condition for the observed interaction-enabled topological pumping. Since the modulation trajectory is symmetric for the first and second half cycles, they should share the same adiabatic condition and therefore here we only analyze the evolution over the first half period, $t\in[0, \pi/\omega]$. We measure the remaining occupation in the initial $\ket{a_0}=\ket{0,0}$ state, $P_{0,0}(t=\pi/\omega)$, with a fixed parameter trajectory and $V=\Delta_c$, but under different modulation frequency $\omega$. As shown in Fig.~\ref{fig3}, in the fast-driving regime the system remains largely frozen in the initial state, indicating a strongly nonadiabatic response. As the modulation is slowed down, the population transfer becomes progressively more efficient, signaling the onset of adiabatic evolution and directional transport in the synthetic lattice. In the slow-driving limit, the system enters a fully adiabatic regime where $P_{0,0}$ saturates to a small residual value. The remaining deviation from the ideal zero occupation is primarily attributed to finite state-preparation fidelity, as discussed above.

To identify the relevant adiabatic scales, we map the dynamics during the first half cycle to an effective two-level state transfer $H=[\Omega(t)\sigma_x + \delta(t)\sigma_z]/2$ with encoding $\ket{\downarrow}=\ket{a_0}$ and $\ket{\uparrow}=\ket{b_0}$. Since the interaction lifts up the on-site energy of $\ket{\uparrow}$ by $V$, the on-site energy difference $\delta(t) = \Delta \cos{(\omega t)}$ with $V=\Delta_c$ in Fig.~\ref{fig3}. Here the Rabi coupling $\Omega(t) = J_0[1+\sin{(\omega t)}]^2$ solely depends on the $J_+$ tone during the first half period. The general adiabatic condition $|\delta \partial_t\Omega - \Omega\partial_t\delta| \ll (\Omega^2 + \delta^2)^{3/2}$ \cite{Rubbmark1981,Zhu1994} then yields two characteristic $\omega$ values, serving as weak and tight adiabatic constraints for the pumping dynamics \cite{SuppMats}. 

In the large $\omega$ limit, adiabatic transitions start working weakly at the boundaries of the time evolution, i.e., at $t=0$ and $t=\pi/\omega$, leading to an upper crossover scale $\omega_u = \Delta^2/(2 J_0)$. Below this value, population transfer becomes efficient. In contrast, near the avoided crossing (two-level resonance point) at $t=\pi/2\omega$, adiabatic following requires a much stronger constraint, yielding a tight scale $\omega_l = 16 J_0^2/\Delta$, below which the evolution becomes fully adiabatic and the quantized pumping behavior emerges. Using the experimental parameters, these two scales correspond to $\omega_u\sim 2\pi\times 9.6~{\rm MHz}$ and $\omega_l\sim 2\pi\times 0.6~{\rm MHz}$, in good agreement with the observed crossover behavior in Fig.~\ref{fig3}.

\begin{figure}[]
	\includegraphics[width=0.48\textwidth]{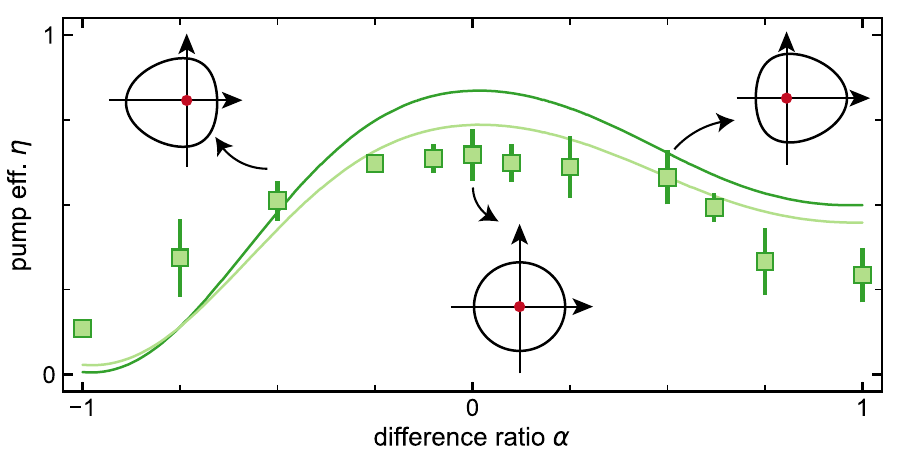}
	\caption{
    \textbf{Response to perturbations of the interaction-enabled topological pumping.} 
    Measured pumping efficiency $\eta$ as a function of the difference ratio $\alpha$ between intra- and inter-cell coupling rates $J_1(t)=(1+\alpha)J_0[1+\sin{(\omega t)}]^2/2$ and $J_2(t)=(1-\alpha)J_0[1-\sin{(\omega t)}]^2/2$. The pumping parameters used are: $V/h=\Delta_c/h = 3.0~{\rm MHz}$, $\Delta/h = 2.4~{\rm MHz}$, $J_0/h = 0.3~{\rm MHz}$, and $\omega/2\pi = 0.3~{\rm MHz}$. The insets respectively correspond to three typical pumping parameter trajectories for $\alpha=0$ and $\pm 0.5$. The solid lines are numerical simulation results with and without the consideration of the initial state preparation inefficiency, respectively. The error bars are standard errors from multiple independent experimental data sets. 
    }
\label{fig4}
\end{figure}

Finally, we test the robustness of the interaction-enabled pumping against controlled distortions of the pumping trajectory. To this end, we introduce an imbalance between intra- and inter-cell couplings, $J_{1,2}(t)=(1\pm\alpha)J_0[1\pm\sin{(\omega t)}]^2/2$, which continuously deforms the modulation loop in parameter space. For $\alpha=-1$, the $J_1$ tone is effectively turned off, and the pumping efficiency vanishes, consistent with a trivial dynamical protocol; see Fig.~\ref{fig4}. For moderate perturbations, $|\alpha| \le 0.5$, the pumping efficiency remains essentially unchanged compared to the unperturbed case. In this regime, the modulation loop continues to enclose the interaction-shifted singularity, and the dynamics remain adiabatic within the accessible driving frequency. However, for large positive $\alpha$, although the trajectory still encloses the singularity, the strongly asymmetric couplings significantly reduce the effective inter-cell transfer rate. This leads to a parametrically reduced adiabatic threshold for the second half-cycle, with $\omega_l \propto (1-\alpha)^2$. As a result, for sufficiently large $\alpha$, e.g., $\omega_l/2\pi\sim 0.04~{\rm MHz}$ for $\alpha=0.75$ (much smaller than experimentally used $\omega/2\pi=0.3~{\rm MHz}$), the system is pushed out of the adiabatic regime and the quantized pumping breaks down \cite{Shih1994,Lindner2017,Privitera2018} even though the underlying topology remains nontrivial.

As summarized in Fig.~\ref{fig4}, the observed pumping is therefore robust against moderate perturbations of the driving protocol, reflecting its topological origin, but its quantization is ultimately constrained by adiabaticity. This distinction between topological protection and dynamical limitations is particularly pronounced in the present interacting setting, where the transport is governed jointly by the interaction-induced motion of the singularity and the multi-step adiabatic passage in the effective two-level subspaces.

In summary, we have experimentally demonstrated interaction-enabled topological pumping of correlated Rydberg electrons in a synthetic lattice. By engineering a bichromatic microwave driving scheme, we directly observe the dipolar-interaction-induced displacement of the topological singularity in parameter space. As the interaction strength is varied, the singularity moves across the pumping trajectory, giving rise to two topological phase transitions and activating quantized directional transport within an intermediate interaction regime. The observed pumping dynamics can be understood as a sequence of interaction-controlled Landau-Zener transfer processes, whose quantization is protected by topology while remaining subject to the adiabaticity requirement of the modulation protocol. We further verify the robustness of the pumping response against perturbations of the driving trajectory as long as the singularity remains enclosed and the evolution stays adiabatic. 

Our work establishes a simple and intuitive picture of how long-range dipolar interactions modify the topology of a driven quantum system through the motion of topological singularities. Beyond the two-particle setting investigated here, the approach could be extended to larger Rydberg arrays, where the interplay between topology and long-range interactions may give rise to genuinely many-body pumping phenomena that have no direct single-particle counterpart. The combination of synthetic dimensions, programmable interactions, and Floquet control also provides a promising route toward studying interaction-controlled Berry curvature engineering \cite{Xiao2010} and topological nature in strongly interacting quantum matter \cite{Jochim-FQH,Wang2025}.

Finally, we note a formal duality between our Rydberg synthetic lattice system and Floquet-driven Hubbard wires in the presence of a potential bias. This duality is exact when the interaction between Rydberg atoms is uniform in space, i.e., for atom pairs and trimers. This duality suggests two exciting directions: that the topological pumping of Hubbard doublons and triplons could be realized in multi-frequency-driven quantum gas microscope experiments, and furthermore that Rydberg synthetic dimensions experiments could be used for few-body explorations of anyonic Hubbard models~\cite{Keilmann2011,Gresch-1,joyce}.

\emph{Acknowledgments. --}
We thank Fabian Heidrich-Meisner for helpful discussions and we thank Tabor Electronics greatly for the use of an arbitrary waveform generator demo unit. This material is based upon work supported by the National Science Foundation under grant No.~1945031 and No.~2438226 and the AFOSR MURI program under agreement number FA9550-22-1-0339.
K.R.A.H.'s work is supported in part by National Science Foundation (PHY-1848304) and W. M. Keck Foundation (Grant No. 995764).

\bibliographystyle{apsrev4-1}
\bibliography{Ryd,TP}

\begin{thebibliography}{62}%
\makeatletter
\providecommand \@ifxundefined [1]{%
 \@ifx{#1\undefined}
}%
\providecommand \@ifnum [1]{%
 \ifnum #1\expandafter \@firstoftwo
 \else \expandafter \@secondoftwo
 \fi
}%
\providecommand \@ifx [1]{%
 \ifx #1\expandafter \@firstoftwo
 \else \expandafter \@secondoftwo
 \fi
}%
\providecommand \natexlab [1]{#1}%
\providecommand \enquote  [1]{``#1''}%
\providecommand \bibnamefont  [1]{#1}%
\providecommand \bibfnamefont [1]{#1}%
\providecommand \citenamefont [1]{#1}%
\providecommand \href@noop [0]{\@secondoftwo}%
\providecommand \href [0]{\begingroup \@sanitize@url \@href}%
\providecommand \@href[1]{\@@startlink{#1}\@@href}%
\providecommand \@@href[1]{\endgroup#1\@@endlink}%
\providecommand \@sanitize@url [0]{\catcode `\\12\catcode `\$12\catcode
  `\&12\catcode `\#12\catcode `\^12\catcode `\_12\catcode `\%12\relax}%
\providecommand \@@startlink[1]{}%
\providecommand \@@endlink[0]{}%
\providecommand \url  [0]{\begingroup\@sanitize@url \@url }%
\providecommand \@url [1]{\endgroup\@href {#1}{\urlprefix }}%
\providecommand \urlprefix  [0]{URL }%
\providecommand \Eprint [0]{\href }%
\providecommand \doibase [0]{http://dx.doi.org/}%
\providecommand \selectlanguage [0]{\@gobble}%
\providecommand \bibinfo  [0]{\@secondoftwo}%
\providecommand \bibfield  [0]{\@secondoftwo}%
\providecommand \translation [1]{[#1]}%
\providecommand \BibitemOpen [0]{}%
\providecommand \bibitemStop [0]{}%
\providecommand \bibitemNoStop [0]{.\EOS\space}%
\providecommand \EOS [0]{\spacefactor3000\relax}%
\providecommand \BibitemShut  [1]{\csname bibitem#1\endcsname}%
\let\auto@bib@innerbib\@empty
\bibitem [{\citenamefont {Haldane}(2017)}]{Haldane2017}%
  \BibitemOpen
  \bibfield  {author} {\bibinfo {author} {\bibfnamefont {F.~D.~M.}\
  \bibnamefont {Haldane}},\ }\href {\doibase 10.1103/RevModPhys.89.040502}
  {\bibfield  {journal} {\bibinfo  {journal} {Rev. Mod. Phys.}\ }\textbf
  {\bibinfo {volume} {89}},\ \bibinfo {pages} {040502} (\bibinfo {year}
  {2017})}\BibitemShut {NoStop}%
\bibitem [{\citenamefont {Nayak}\ \emph {et~al.}(2008)\citenamefont {Nayak},
  \citenamefont {Simon}, \citenamefont {Stern}, \citenamefont {Freedman},\ and\
  \citenamefont {Das~Sarma}}]{Chetan2008}%
  \BibitemOpen
  \bibfield  {author} {\bibinfo {author} {\bibfnamefont {C.}~\bibnamefont
  {Nayak}}, \bibinfo {author} {\bibfnamefont {S.~H.}\ \bibnamefont {Simon}},
  \bibinfo {author} {\bibfnamefont {A.}~\bibnamefont {Stern}}, \bibinfo
  {author} {\bibfnamefont {M.}~\bibnamefont {Freedman}}, \ and\ \bibinfo
  {author} {\bibfnamefont {S.}~\bibnamefont {Das~Sarma}},\ }\href {\doibase
  10.1103/RevModPhys.80.1083} {\bibfield  {journal} {\bibinfo  {journal} {Rev.
  Mod. Phys.}\ }\textbf {\bibinfo {volume} {80}},\ \bibinfo {pages} {1083}
  (\bibinfo {year} {2008})}\BibitemShut {NoStop}%
\bibitem [{\citenamefont {Wen}(2004)}]{Wen2004}%
  \BibitemOpen
  \bibfield  {author} {\bibinfo {author} {\bibfnamefont {X.-G.}\ \bibnamefont
  {Wen}},\ }\href@noop {} {\emph {\bibinfo {title} {Quantum field theory of
  many-body systems}}}\ (\bibinfo  {publisher} {Oxford university press},\
  \bibinfo {year} {2004})\BibitemShut {NoStop}%
\bibitem [{\citenamefont {Cooper}\ \emph {et~al.}(2019)\citenamefont {Cooper},
  \citenamefont {Dalibard},\ and\ \citenamefont {Spielman}}]{Cooper2019}%
  \BibitemOpen
  \bibfield  {author} {\bibinfo {author} {\bibfnamefont {N.~R.}\ \bibnamefont
  {Cooper}}, \bibinfo {author} {\bibfnamefont {J.}~\bibnamefont {Dalibard}}, \
  and\ \bibinfo {author} {\bibfnamefont {I.~B.}\ \bibnamefont {Spielman}},\
  }\href {\doibase 10.1103/RevModPhys.91.015005} {\bibfield  {journal}
  {\bibinfo  {journal} {Rev. Mod. Phys.}\ }\textbf {\bibinfo {volume} {91}},\
  \bibinfo {pages} {015005} (\bibinfo {year} {2019})}\BibitemShut {NoStop}%
\bibitem [{\citenamefont
  {Thouless}(1983)}]{thoulessQuantizationParticleTransport1983}%
  \BibitemOpen
  \bibfield  {author} {\bibinfo {author} {\bibfnamefont {D.~J.}\ \bibnamefont
  {Thouless}},\ }\href {\doibase 10.1103/PhysRevB.27.6083} {\bibfield
  {journal} {\bibinfo  {journal} {Physical Review B}\ }\textbf {\bibinfo
  {volume} {27}},\ \bibinfo {pages} {6083} (\bibinfo {year}
  {1983})}\BibitemShut {NoStop}%
\bibitem [{\citenamefont {Niu}\ and\ \citenamefont {Thouless}(1984)}]{Niu1984}%
  \BibitemOpen
  \bibfield  {author} {\bibinfo {author} {\bibfnamefont {Q.}~\bibnamefont
  {Niu}}\ and\ \bibinfo {author} {\bibfnamefont {D.}~\bibnamefont {Thouless}},\
  }\href {https://iopscience.iop.org/article/10.1088/0305-4470/17/12/016}
  {\bibfield  {journal} {\bibinfo  {journal} {J. Phys. A: Math. Gen.}\ }\textbf
  {\bibinfo {volume} {17}},\ \bibinfo {pages} {2453} (\bibinfo {year}
  {1984})}\BibitemShut {NoStop}%
\bibitem [{\citenamefont {Citro}\ and\ \citenamefont
  {Aidelsburger}(2023)}]{citroThoulessPumpingTopology2023}%
  \BibitemOpen
  \bibfield  {author} {\bibinfo {author} {\bibfnamefont {R.}~\bibnamefont
  {Citro}}\ and\ \bibinfo {author} {\bibfnamefont {M.}~\bibnamefont
  {Aidelsburger}},\ }\href {\doibase 10.1038/s42254-022-00545-0} {\bibfield
  {journal} {\bibinfo  {journal} {Nature Reviews Physics}\ }\textbf {\bibinfo
  {volume} {5}},\ \bibinfo {pages} {87} (\bibinfo {year} {2023})}\BibitemShut
  {NoStop}%
\bibitem [{\citenamefont {Lohse}\ \emph {et~al.}(2016)\citenamefont {Lohse},
  \citenamefont {Schweizer}, \citenamefont {Zilberberg}, \citenamefont
  {Aidelsburger},\ and\ \citenamefont {Bloch}}]{lohseThoulessQuantumPump2016}%
  \BibitemOpen
  \bibfield  {author} {\bibinfo {author} {\bibfnamefont {M.}~\bibnamefont
  {Lohse}}, \bibinfo {author} {\bibfnamefont {C.}~\bibnamefont {Schweizer}},
  \bibinfo {author} {\bibfnamefont {O.}~\bibnamefont {Zilberberg}}, \bibinfo
  {author} {\bibfnamefont {M.}~\bibnamefont {Aidelsburger}}, \ and\ \bibinfo
  {author} {\bibfnamefont {I.}~\bibnamefont {Bloch}},\ }\href {\doibase
  10.1038/nphys3584} {\bibfield  {journal} {\bibinfo  {journal} {Nature
  Physics}\ }\textbf {\bibinfo {volume} {12}},\ \bibinfo {pages} {350}
  (\bibinfo {year} {2016})}\BibitemShut {NoStop}%
\bibitem [{\citenamefont {Nakajima}\ \emph {et~al.}(2016)\citenamefont
  {Nakajima}, \citenamefont {Tomita}, \citenamefont {Taie}, \citenamefont
  {Ichinose}, \citenamefont {Ozawa}, \citenamefont {Wang}, \citenamefont
  {Troyer},\ and\ \citenamefont
  {Takahashi}}]{nakajimaTopologicalThoulessPumping2016}%
  \BibitemOpen
  \bibfield  {author} {\bibinfo {author} {\bibfnamefont {S.}~\bibnamefont
  {Nakajima}}, \bibinfo {author} {\bibfnamefont {T.}~\bibnamefont {Tomita}},
  \bibinfo {author} {\bibfnamefont {S.}~\bibnamefont {Taie}}, \bibinfo {author}
  {\bibfnamefont {T.}~\bibnamefont {Ichinose}}, \bibinfo {author}
  {\bibfnamefont {H.}~\bibnamefont {Ozawa}}, \bibinfo {author} {\bibfnamefont
  {L.}~\bibnamefont {Wang}}, \bibinfo {author} {\bibfnamefont {M.}~\bibnamefont
  {Troyer}}, \ and\ \bibinfo {author} {\bibfnamefont {Y.}~\bibnamefont
  {Takahashi}},\ }\href {\doibase 10.1038/nphys3622} {\bibfield  {journal}
  {\bibinfo  {journal} {Nature Physics}\ }\textbf {\bibinfo {volume} {12}},\
  \bibinfo {pages} {296} (\bibinfo {year} {2016})}\BibitemShut {NoStop}%
\bibitem [{\citenamefont {Lu}\ \emph {et~al.}(2016)\citenamefont {Lu},
  \citenamefont {Schemmer}, \citenamefont {Aycock}, \citenamefont {Genkina},
  \citenamefont {Sugawa},\ and\ \citenamefont {Spielman}}]{Lu2016}%
  \BibitemOpen
  \bibfield  {author} {\bibinfo {author} {\bibfnamefont {H.-I.}\ \bibnamefont
  {Lu}}, \bibinfo {author} {\bibfnamefont {M.}~\bibnamefont {Schemmer}},
  \bibinfo {author} {\bibfnamefont {L.~M.}\ \bibnamefont {Aycock}}, \bibinfo
  {author} {\bibfnamefont {D.}~\bibnamefont {Genkina}}, \bibinfo {author}
  {\bibfnamefont {S.}~\bibnamefont {Sugawa}}, \ and\ \bibinfo {author}
  {\bibfnamefont {I.~B.}\ \bibnamefont {Spielman}},\ }\href {\doibase
  10.1103/PhysRevLett.116.200402} {\bibfield  {journal} {\bibinfo  {journal}
  {Phys. Rev. Lett.}\ }\textbf {\bibinfo {volume} {116}},\ \bibinfo {pages}
  {200402} (\bibinfo {year} {2016})}\BibitemShut {NoStop}%
\bibitem [{\citenamefont {Laughlin}(1983)}]{Laughlin1983}%
  \BibitemOpen
  \bibfield  {author} {\bibinfo {author} {\bibfnamefont {R.~B.}\ \bibnamefont
  {Laughlin}},\ }\href {\doibase 10.1103/PhysRevLett.50.1395} {\bibfield
  {journal} {\bibinfo  {journal} {Phys. Rev. Lett.}\ }\textbf {\bibinfo
  {volume} {50}},\ \bibinfo {pages} {1395} (\bibinfo {year}
  {1983})}\BibitemShut {NoStop}%
\bibitem [{\citenamefont {Hafezi}\ \emph {et~al.}(2007)\citenamefont {Hafezi},
  \citenamefont {S\o{}rensen}, \citenamefont {Demler},\ and\ \citenamefont
  {Lukin}}]{Hafezi2007}%
  \BibitemOpen
  \bibfield  {author} {\bibinfo {author} {\bibfnamefont {M.}~\bibnamefont
  {Hafezi}}, \bibinfo {author} {\bibfnamefont {A.~S.}\ \bibnamefont
  {S\o{}rensen}}, \bibinfo {author} {\bibfnamefont {E.}~\bibnamefont {Demler}},
  \ and\ \bibinfo {author} {\bibfnamefont {M.~D.}\ \bibnamefont {Lukin}},\
  }\href {\doibase 10.1103/PhysRevA.76.023613} {\bibfield  {journal} {\bibinfo
  {journal} {Phys. Rev. A}\ }\textbf {\bibinfo {volume} {76}},\ \bibinfo
  {pages} {023613} (\bibinfo {year} {2007})}\BibitemShut {NoStop}%
\bibitem [{\citenamefont {Clark}\ \emph {et~al.}(2020)\citenamefont {Clark},
  \citenamefont {Schine}, \citenamefont {Baum}, \citenamefont {Jia},\ and\
  \citenamefont {Simon}}]{Clark2020}%
  \BibitemOpen
  \bibfield  {author} {\bibinfo {author} {\bibfnamefont {L.~W.}\ \bibnamefont
  {Clark}}, \bibinfo {author} {\bibfnamefont {N.}~\bibnamefont {Schine}},
  \bibinfo {author} {\bibfnamefont {C.}~\bibnamefont {Baum}}, \bibinfo {author}
  {\bibfnamefont {N.}~\bibnamefont {Jia}}, \ and\ \bibinfo {author}
  {\bibfnamefont {J.}~\bibnamefont {Simon}},\ }\href
  {https://www.nature.com/articles/s41586-020-2318-5} {\bibfield  {journal}
  {\bibinfo  {journal} {Nature}\ }\textbf {\bibinfo {volume} {582}},\ \bibinfo
  {pages} {41} (\bibinfo {year} {2020})}\BibitemShut {NoStop}%
\bibitem [{\citenamefont {L{\'e}onard}\ \emph {et~al.}(2023)\citenamefont
  {L{\'e}onard}, \citenamefont {Kim}, \citenamefont {Kwan}, \citenamefont
  {Segura}, \citenamefont {Grusdt}, \citenamefont {Repellin}, \citenamefont
  {Goldman},\ and\ \citenamefont {Greiner}}]{Leonard2023}%
  \BibitemOpen
  \bibfield  {author} {\bibinfo {author} {\bibfnamefont {J.}~\bibnamefont
  {L{\'e}onard}}, \bibinfo {author} {\bibfnamefont {S.}~\bibnamefont {Kim}},
  \bibinfo {author} {\bibfnamefont {J.}~\bibnamefont {Kwan}}, \bibinfo {author}
  {\bibfnamefont {P.}~\bibnamefont {Segura}}, \bibinfo {author} {\bibfnamefont
  {F.}~\bibnamefont {Grusdt}}, \bibinfo {author} {\bibfnamefont
  {C.}~\bibnamefont {Repellin}}, \bibinfo {author} {\bibfnamefont
  {N.}~\bibnamefont {Goldman}}, \ and\ \bibinfo {author} {\bibfnamefont
  {M.}~\bibnamefont {Greiner}},\ }\href
  {https://www.nature.com/articles/s41586-023-06122-4} {\bibfield  {journal}
  {\bibinfo  {journal} {Nature}\ }\textbf {\bibinfo {volume} {619}},\ \bibinfo
  {pages} {495} (\bibinfo {year} {2023})}\BibitemShut {NoStop}%
\bibitem [{\citenamefont {Bohm}\ \emph {et~al.}(2026)\citenamefont {Bohm},
  \citenamefont {Gerlitz}, \citenamefont {J\"org},\ and\ \citenamefont
  {Fleischhauer}}]{Bohm2026}%
  \BibitemOpen
  \bibfield  {author} {\bibinfo {author} {\bibfnamefont {J.}~\bibnamefont
  {Bohm}}, \bibinfo {author} {\bibfnamefont {H.}~\bibnamefont {Gerlitz}},
  \bibinfo {author} {\bibfnamefont {C.}~\bibnamefont {J\"org}}, \ and\ \bibinfo
  {author} {\bibfnamefont {M.}~\bibnamefont {Fleischhauer}},\ }\href {\doibase
  10.1103/hhgy-9rd4} {\bibfield  {journal} {\bibinfo  {journal} {Phys. Rev. X}\
  }\textbf {\bibinfo {volume} {16}},\ \bibinfo {pages} {011038} (\bibinfo
  {year} {2026})}\BibitemShut {NoStop}%
\bibitem [{\citenamefont {Kuno}\ and\ \citenamefont
  {Hatsugai}(2020)}]{Kuno2020}%
  \BibitemOpen
  \bibfield  {author} {\bibinfo {author} {\bibfnamefont {Y.}~\bibnamefont
  {Kuno}}\ and\ \bibinfo {author} {\bibfnamefont {Y.}~\bibnamefont
  {Hatsugai}},\ }\href {\doibase 10.1103/PhysRevResearch.2.042024} {\bibfield
  {journal} {\bibinfo  {journal} {Phys. Rev. Res.}\ }\textbf {\bibinfo {volume}
  {2}},\ \bibinfo {pages} {042024(R)} (\bibinfo {year} {2020})}\BibitemShut
  {NoStop}%
\bibitem [{\citenamefont {Kuno}\ and\ \citenamefont
  {Hatsugai}(2024)}]{kuno2024topologicaldomainwallpumpmathbbz2}%
  \BibitemOpen
  \bibfield  {author} {\bibinfo {author} {\bibfnamefont {Y.}~\bibnamefont
  {Kuno}}\ and\ \bibinfo {author} {\bibfnamefont {Y.}~\bibnamefont
  {Hatsugai}},\ }\href {https://arxiv.org/abs/2411.17219} {\enquote {\bibinfo
  {title} {Topological domain-wall pump with $\mathbb{Z}_2$ spontaneous
  symmetry breaking},}\ } (\bibinfo {year} {2024}),\ \Eprint
  {http://arxiv.org/abs/2411.17219} {arXiv:2411.17219 [cond-mat.str-el]}
  \BibitemShut {NoStop}%
\bibitem [{\citenamefont {Yan}\ and\ \citenamefont
  {Zhou}(2018)}]{yanYangMonopolesEmergent2018}%
  \BibitemOpen
  \bibfield  {author} {\bibinfo {author} {\bibfnamefont {Y.}~\bibnamefont
  {Yan}}\ and\ \bibinfo {author} {\bibfnamefont {Q.}~\bibnamefont {Zhou}},\
  }\href {\doibase 10.1103/PhysRevLett.120.235302} {\bibfield  {journal}
  {\bibinfo  {journal} {Physical Review Letters}\ }\textbf {\bibinfo {volume}
  {120}},\ \bibinfo {pages} {235302} (\bibinfo {year} {2018})}\BibitemShut
  {NoStop}%
\bibitem [{\citenamefont {Lam}\ and\ \citenamefont {Yan}(2024)}]{Lam2024}%
  \BibitemOpen
  \bibfield  {author} {\bibinfo {author} {\bibfnamefont {H.}~\bibnamefont
  {Lam}}\ and\ \bibinfo {author} {\bibfnamefont {Y.}~\bibnamefont {Yan}},\
  }\href {https://arxiv.org/abs/2405.10291} {\enquote {\bibinfo {title}
  {Interaction induced splitting of dirac monopoles in the topological thouless
  pumping of strongly interacting bosons and su($n$) fermions},}\ } (\bibinfo
  {year} {2024}),\ \Eprint {http://arxiv.org/abs/2405.10291} {arXiv:2405.10291
  [cond-mat.quant-gas]} \BibitemShut {NoStop}%
\bibitem [{\citenamefont {Mostaan}\ \emph {et~al.}(2022)\citenamefont
  {Mostaan}, \citenamefont {Grusdt},\ and\ \citenamefont
  {Goldman}}]{Mostaan2022}%
  \BibitemOpen
  \bibfield  {author} {\bibinfo {author} {\bibfnamefont {N.}~\bibnamefont
  {Mostaan}}, \bibinfo {author} {\bibfnamefont {F.}~\bibnamefont {Grusdt}}, \
  and\ \bibinfo {author} {\bibfnamefont {N.}~\bibnamefont {Goldman}},\ }\href
  {\doibase 10.1038/s41467-022-33478-4} {\bibfield  {journal} {\bibinfo
  {journal} {Nature Communications}\ }\textbf {\bibinfo {volume} {13}},\
  \bibinfo {pages} {5997} (\bibinfo {year} {2022})}\BibitemShut {NoStop}%
\bibitem [{\citenamefont {Zeng}\ \emph {et~al.}(2015)\citenamefont {Zeng},
  \citenamefont {Wang},\ and\ \citenamefont {Zhai}}]{Zeng2015}%
  \BibitemOpen
  \bibfield  {author} {\bibinfo {author} {\bibfnamefont {T.-S.}\ \bibnamefont
  {Zeng}}, \bibinfo {author} {\bibfnamefont {C.}~\bibnamefont {Wang}}, \ and\
  \bibinfo {author} {\bibfnamefont {H.}~\bibnamefont {Zhai}},\ }\href {\doibase
  10.1103/PhysRevLett.115.095302} {\bibfield  {journal} {\bibinfo  {journal}
  {Phys. Rev. Lett.}\ }\textbf {\bibinfo {volume} {115}},\ \bibinfo {pages}
  {095302} (\bibinfo {year} {2015})}\BibitemShut {NoStop}%
\bibitem [{\citenamefont {Zeng}\ \emph {et~al.}(2016)\citenamefont {Zeng},
  \citenamefont {Zhu},\ and\ \citenamefont {Sheng}}]{Zeng2016}%
  \BibitemOpen
  \bibfield  {author} {\bibinfo {author} {\bibfnamefont {T.-S.}\ \bibnamefont
  {Zeng}}, \bibinfo {author} {\bibfnamefont {W.}~\bibnamefont {Zhu}}, \ and\
  \bibinfo {author} {\bibfnamefont {D.~N.}\ \bibnamefont {Sheng}},\ }\href
  {\doibase 10.1103/PhysRevB.94.235139} {\bibfield  {journal} {\bibinfo
  {journal} {Phys. Rev. B}\ }\textbf {\bibinfo {volume} {94}},\ \bibinfo
  {pages} {235139} (\bibinfo {year} {2016})}\BibitemShut {NoStop}%
\bibitem [{\citenamefont {Berg}\ \emph {et~al.}(2011)\citenamefont {Berg},
  \citenamefont {Levin},\ and\ \citenamefont {Altman}}]{Berg2011}%
  \BibitemOpen
  \bibfield  {author} {\bibinfo {author} {\bibfnamefont {E.}~\bibnamefont
  {Berg}}, \bibinfo {author} {\bibfnamefont {M.}~\bibnamefont {Levin}}, \ and\
  \bibinfo {author} {\bibfnamefont {E.}~\bibnamefont {Altman}},\ }\href
  {\doibase 10.1103/PhysRevLett.106.110405} {\bibfield  {journal} {\bibinfo
  {journal} {Phys. Rev. Lett.}\ }\textbf {\bibinfo {volume} {106}},\ \bibinfo
  {pages} {110405} (\bibinfo {year} {2011})}\BibitemShut {NoStop}%
\bibitem [{\citenamefont {Taddia}\ \emph {et~al.}(2017)\citenamefont {Taddia},
  \citenamefont {Cornfeld}, \citenamefont {Rossini}, \citenamefont {Mazza},
  \citenamefont {Sela},\ and\ \citenamefont {Fazio}}]{Taddia2017}%
  \BibitemOpen
  \bibfield  {author} {\bibinfo {author} {\bibfnamefont {L.}~\bibnamefont
  {Taddia}}, \bibinfo {author} {\bibfnamefont {E.}~\bibnamefont {Cornfeld}},
  \bibinfo {author} {\bibfnamefont {D.}~\bibnamefont {Rossini}}, \bibinfo
  {author} {\bibfnamefont {L.}~\bibnamefont {Mazza}}, \bibinfo {author}
  {\bibfnamefont {E.}~\bibnamefont {Sela}}, \ and\ \bibinfo {author}
  {\bibfnamefont {R.}~\bibnamefont {Fazio}},\ }\href {\doibase
  10.1103/PhysRevLett.118.230402} {\bibfield  {journal} {\bibinfo  {journal}
  {Phys. Rev. Lett.}\ }\textbf {\bibinfo {volume} {118}},\ \bibinfo {pages}
  {230402} (\bibinfo {year} {2017})}\BibitemShut {NoStop}%
\bibitem [{\citenamefont {Nakagawa}\ \emph {et~al.}(2018)\citenamefont
  {Nakagawa}, \citenamefont {Yoshida}, \citenamefont {Peters},\ and\
  \citenamefont {Kawakami}}]{Nakagawa2018}%
  \BibitemOpen
  \bibfield  {author} {\bibinfo {author} {\bibfnamefont {M.}~\bibnamefont
  {Nakagawa}}, \bibinfo {author} {\bibfnamefont {T.}~\bibnamefont {Yoshida}},
  \bibinfo {author} {\bibfnamefont {R.}~\bibnamefont {Peters}}, \ and\ \bibinfo
  {author} {\bibfnamefont {N.}~\bibnamefont {Kawakami}},\ }\href {\doibase
  10.1103/PhysRevB.98.115147} {\bibfield  {journal} {\bibinfo  {journal} {Phys.
  Rev. B}\ }\textbf {\bibinfo {volume} {98}},\ \bibinfo {pages} {115147}
  (\bibinfo {year} {2018})}\BibitemShut {NoStop}%
\bibitem [{\citenamefont {Unanyan}\ \emph {et~al.}(2020)\citenamefont
  {Unanyan}, \citenamefont {Kiefer-Emmanouilidis},\ and\ \citenamefont
  {Fleischhauer}}]{Unanyan2020}%
  \BibitemOpen
  \bibfield  {author} {\bibinfo {author} {\bibfnamefont {R.}~\bibnamefont
  {Unanyan}}, \bibinfo {author} {\bibfnamefont {M.}~\bibnamefont
  {Kiefer-Emmanouilidis}}, \ and\ \bibinfo {author} {\bibfnamefont
  {M.}~\bibnamefont {Fleischhauer}},\ }\href {\doibase
  10.1103/PhysRevLett.125.215701} {\bibfield  {journal} {\bibinfo  {journal}
  {Phys. Rev. Lett.}\ }\textbf {\bibinfo {volume} {125}},\ \bibinfo {pages}
  {215701} (\bibinfo {year} {2020})}\BibitemShut {NoStop}%
\bibitem [{\citenamefont {Chen}\ \emph {et~al.}(2020)\citenamefont {Chen},
  \citenamefont {Zhang}, \citenamefont {Zhang},\ and\ \citenamefont
  {Zhu}}]{Chenyl2020}%
  \BibitemOpen
  \bibfield  {author} {\bibinfo {author} {\bibfnamefont {Y.-L.}\ \bibnamefont
  {Chen}}, \bibinfo {author} {\bibfnamefont {G.-Q.}\ \bibnamefont {Zhang}},
  \bibinfo {author} {\bibfnamefont {D.-W.}\ \bibnamefont {Zhang}}, \ and\
  \bibinfo {author} {\bibfnamefont {S.-L.}\ \bibnamefont {Zhu}},\ }\href
  {\doibase 10.1103/PhysRevA.101.013627} {\bibfield  {journal} {\bibinfo
  {journal} {Phys. Rev. A}\ }\textbf {\bibinfo {volume} {101}},\ \bibinfo
  {pages} {013627} (\bibinfo {year} {2020})}\BibitemShut {NoStop}%
\bibitem [{\citenamefont {Fu}\ \emph {et~al.}(2022{\natexlab{a}})\citenamefont
  {Fu}, \citenamefont {Wang}, \citenamefont {Kartashov}, \citenamefont
  {Konotop},\ and\ \citenamefont {Ye}}]{Fu2022a}%
  \BibitemOpen
  \bibfield  {author} {\bibinfo {author} {\bibfnamefont {Q.}~\bibnamefont
  {Fu}}, \bibinfo {author} {\bibfnamefont {P.}~\bibnamefont {Wang}}, \bibinfo
  {author} {\bibfnamefont {Y.~V.}\ \bibnamefont {Kartashov}}, \bibinfo {author}
  {\bibfnamefont {V.~V.}\ \bibnamefont {Konotop}}, \ and\ \bibinfo {author}
  {\bibfnamefont {F.}~\bibnamefont {Ye}},\ }\href {\doibase
  10.1103/PhysRevLett.128.154101} {\bibfield  {journal} {\bibinfo  {journal}
  {Phys. Rev. Lett.}\ }\textbf {\bibinfo {volume} {128}},\ \bibinfo {pages}
  {154101} (\bibinfo {year} {2022}{\natexlab{a}})}\BibitemShut {NoStop}%
\bibitem [{\citenamefont {Fu}\ \emph {et~al.}(2022{\natexlab{b}})\citenamefont
  {Fu}, \citenamefont {Wang}, \citenamefont {Kartashov}, \citenamefont
  {Konotop},\ and\ \citenamefont {Ye}}]{Fu2022b}%
  \BibitemOpen
  \bibfield  {author} {\bibinfo {author} {\bibfnamefont {Q.}~\bibnamefont
  {Fu}}, \bibinfo {author} {\bibfnamefont {P.}~\bibnamefont {Wang}}, \bibinfo
  {author} {\bibfnamefont {Y.~V.}\ \bibnamefont {Kartashov}}, \bibinfo {author}
  {\bibfnamefont {V.~V.}\ \bibnamefont {Konotop}}, \ and\ \bibinfo {author}
  {\bibfnamefont {F.}~\bibnamefont {Ye}},\ }\href {\doibase
  10.1103/PhysRevLett.129.183901} {\bibfield  {journal} {\bibinfo  {journal}
  {Phys. Rev. Lett.}\ }\textbf {\bibinfo {volume} {129}},\ \bibinfo {pages}
  {183901} (\bibinfo {year} {2022}{\natexlab{b}})}\BibitemShut {NoStop}%
\bibitem [{\citenamefont {Ke}\ \emph {et~al.}(2017)\citenamefont {Ke},
  \citenamefont {Qin}, \citenamefont {Kivshar},\ and\ \citenamefont
  {Lee}}]{Ke2017}%
  \BibitemOpen
  \bibfield  {author} {\bibinfo {author} {\bibfnamefont {Y.}~\bibnamefont
  {Ke}}, \bibinfo {author} {\bibfnamefont {X.}~\bibnamefont {Qin}}, \bibinfo
  {author} {\bibfnamefont {Y.~S.}\ \bibnamefont {Kivshar}}, \ and\ \bibinfo
  {author} {\bibfnamefont {C.}~\bibnamefont {Lee}},\ }\href {\doibase
  10.1103/PhysRevA.95.063630} {\bibfield  {journal} {\bibinfo  {journal} {Phys.
  Rev. A}\ }\textbf {\bibinfo {volume} {95}},\ \bibinfo {pages} {063630}
  (\bibinfo {year} {2017})}\BibitemShut {NoStop}%
\bibitem [{\citenamefont {Huang}\ \emph {et~al.}(2024)\citenamefont {Huang},
  \citenamefont {Ke}, \citenamefont {Zhong}, \citenamefont {Kivshar},\ and\
  \citenamefont {Lee}}]{Huang2024}%
  \BibitemOpen
  \bibfield  {author} {\bibinfo {author} {\bibfnamefont {B.}~\bibnamefont
  {Huang}}, \bibinfo {author} {\bibfnamefont {Y.}~\bibnamefont {Ke}}, \bibinfo
  {author} {\bibfnamefont {H.}~\bibnamefont {Zhong}}, \bibinfo {author}
  {\bibfnamefont {Y.~S.}\ \bibnamefont {Kivshar}}, \ and\ \bibinfo {author}
  {\bibfnamefont {C.}~\bibnamefont {Lee}},\ }\href {\doibase
  10.1103/PhysRevLett.133.140202} {\bibfield  {journal} {\bibinfo  {journal}
  {Phys. Rev. Lett.}\ }\textbf {\bibinfo {volume} {133}},\ \bibinfo {pages}
  {140202} (\bibinfo {year} {2024})}\BibitemShut {NoStop}%
\bibitem [{\citenamefont {Wu}\ \emph {et~al.}(2025)\citenamefont {Wu},
  \citenamefont {Zuo}, \citenamefont {Zhu}, \citenamefont {Yuan}, \citenamefont
  {Mao}, \citenamefont {Zeng}, \citenamefont {Jiang}, \citenamefont {Chen},
  \citenamefont {Pan}, \citenamefont {Zheng},\ and\ \citenamefont
  {Dai}}]{Wu2025}%
  \BibitemOpen
  \bibfield  {author} {\bibinfo {author} {\bibfnamefont {F.-F.}\ \bibnamefont
  {Wu}}, \bibinfo {author} {\bibfnamefont {X.-D.}\ \bibnamefont {Zuo}},
  \bibinfo {author} {\bibfnamefont {Q.-Q.}\ \bibnamefont {Zhu}}, \bibinfo
  {author} {\bibfnamefont {T.}~\bibnamefont {Yuan}}, \bibinfo {author}
  {\bibfnamefont {Y.-Y.}\ \bibnamefont {Mao}}, \bibinfo {author} {\bibfnamefont
  {C.}~\bibnamefont {Zeng}}, \bibinfo {author} {\bibfnamefont {Y.}~\bibnamefont
  {Jiang}}, \bibinfo {author} {\bibfnamefont {Y.-A.}\ \bibnamefont {Chen}},
  \bibinfo {author} {\bibfnamefont {J.-W.}\ \bibnamefont {Pan}}, \bibinfo
  {author} {\bibfnamefont {W.}~\bibnamefont {Zheng}}, \ and\ \bibinfo {author}
  {\bibfnamefont {H.-N.}\ \bibnamefont {Dai}},\ }\href
  {https://arxiv.org/abs/2506.08502} {\enquote {\bibinfo {title} {Topological
  invariants in nonlinear thouless pumping of solitons},}\ } (\bibinfo {year}
  {2025}),\ \Eprint {http://arxiv.org/abs/2506.08502} {arXiv:2506.08502
  [physics.atom-ph]} \BibitemShut {NoStop}%
\bibitem [{\citenamefont {J\"urgensen}\ and\ \citenamefont
  {Rechtsman}(2022)}]{Jurgensen2022}%
  \BibitemOpen
  \bibfield  {author} {\bibinfo {author} {\bibfnamefont {M.}~\bibnamefont
  {J\"urgensen}}\ and\ \bibinfo {author} {\bibfnamefont {M.~C.}\ \bibnamefont
  {Rechtsman}},\ }\href {\doibase 10.1103/PhysRevLett.128.113901} {\bibfield
  {journal} {\bibinfo  {journal} {Phys. Rev. Lett.}\ }\textbf {\bibinfo
  {volume} {128}},\ \bibinfo {pages} {113901} (\bibinfo {year}
  {2022})}\BibitemShut {NoStop}%
\bibitem [{\citenamefont {J\"urgensen}\ \emph {et~al.}(2025)\citenamefont
  {J\"urgensen}, \citenamefont {Steiner}, \citenamefont {Refael},\ and\
  \citenamefont {Rechtsman}}]{Jurgensen2025}%
  \BibitemOpen
  \bibfield  {author} {\bibinfo {author} {\bibfnamefont {M.}~\bibnamefont
  {J\"urgensen}}, \bibinfo {author} {\bibfnamefont {J.}~\bibnamefont
  {Steiner}}, \bibinfo {author} {\bibfnamefont {G.}~\bibnamefont {Refael}}, \
  and\ \bibinfo {author} {\bibfnamefont {M.~C.}\ \bibnamefont {Rechtsman}},\
  }\href {\doibase 10.1103/4d5s-n4gn} {\bibfield  {journal} {\bibinfo
  {journal} {Phys. Rev. Lett.}\ }\textbf {\bibinfo {volume} {135}},\ \bibinfo
  {pages} {166601} (\bibinfo {year} {2025})}\BibitemShut {NoStop}%
\bibitem [{\citenamefont {Tao}\ \emph {et~al.}(2025)\citenamefont {Tao},
  \citenamefont {Zhang},\ and\ \citenamefont {Xu}}]{Tao2025}%
  \BibitemOpen
  \bibfield  {author} {\bibinfo {author} {\bibfnamefont {Y.-L.}\ \bibnamefont
  {Tao}}, \bibinfo {author} {\bibfnamefont {Y.}~\bibnamefont {Zhang}}, \ and\
  \bibinfo {author} {\bibfnamefont {Y.}~\bibnamefont {Xu}},\ }\href {\doibase
  10.1103/96f5-qszj} {\bibfield  {journal} {\bibinfo  {journal} {Phys. Rev.
  Lett.}\ }\textbf {\bibinfo {volume} {135}},\ \bibinfo {pages} {097202}
  (\bibinfo {year} {2025})}\BibitemShut {NoStop}%
\bibitem [{\citenamefont {J{\"u}rgensen}\ \emph {et~al.}(2021)\citenamefont
  {J{\"u}rgensen}, \citenamefont {Mukherjee},\ and\ \citenamefont
  {Rechtsman}}]{jurgensenQuantizedNonlinearThouless2021}%
  \BibitemOpen
  \bibfield  {author} {\bibinfo {author} {\bibfnamefont {M.}~\bibnamefont
  {J{\"u}rgensen}}, \bibinfo {author} {\bibfnamefont {S.}~\bibnamefont
  {Mukherjee}}, \ and\ \bibinfo {author} {\bibfnamefont {M.~C.}\ \bibnamefont
  {Rechtsman}},\ }\href {\doibase 10.1038/s41586-021-03688-9} {\bibfield
  {journal} {\bibinfo  {journal} {Nature}\ }\textbf {\bibinfo {volume} {596}},\
  \bibinfo {pages} {63} (\bibinfo {year} {2021})}\BibitemShut {NoStop}%
\bibitem [{\citenamefont {J{\"u}rgensen}\ \emph {et~al.}(2023)\citenamefont
  {J{\"u}rgensen}, \citenamefont {Mukherjee}, \citenamefont {J{\"o}rg},\ and\
  \citenamefont {Rechtsman}}]{jurgensenQuantizedFractionalThouless2023}%
  \BibitemOpen
  \bibfield  {author} {\bibinfo {author} {\bibfnamefont {M.}~\bibnamefont
  {J{\"u}rgensen}}, \bibinfo {author} {\bibfnamefont {S.}~\bibnamefont
  {Mukherjee}}, \bibinfo {author} {\bibfnamefont {C.}~\bibnamefont {J{\"o}rg}},
  \ and\ \bibinfo {author} {\bibfnamefont {M.~C.}\ \bibnamefont {Rechtsman}},\
  }\href {\doibase 10.1038/s41567-022-01871-x} {\bibfield  {journal} {\bibinfo
  {journal} {Nature Physics}\ }\textbf {\bibinfo {volume} {19}},\ \bibinfo
  {pages} {420} (\bibinfo {year} {2023})}\BibitemShut {NoStop}%
\bibitem [{\citenamefont {Chaudhari}\ \emph {et~al.}(2025)\citenamefont
  {Chaudhari}, \citenamefont {Jürgensen},\ and\ \citenamefont
  {Rechtsman}}]{Chaudhari2025}%
  \BibitemOpen
  \bibfield  {author} {\bibinfo {author} {\bibfnamefont {A.~P.}\ \bibnamefont
  {Chaudhari}}, \bibinfo {author} {\bibfnamefont {M.}~\bibnamefont
  {Jürgensen}}, \ and\ \bibinfo {author} {\bibfnamefont {M.~C.}\ \bibnamefont
  {Rechtsman}},\ }\href {https://arxiv.org/abs/2512.11394} {\enquote {\bibinfo
  {title} {Quantized pumping in disordered nonlinear thouless pumps},}\ }
  (\bibinfo {year} {2025}),\ \Eprint {http://arxiv.org/abs/2512.11394}
  {arXiv:2512.11394 [cond-mat.mes-hall]} \BibitemShut {NoStop}%
\bibitem [{\citenamefont {Walter}\ \emph {et~al.}(2023)\citenamefont {Walter},
  \citenamefont {Zhu}, \citenamefont {G{\"a}chter}, \citenamefont {Minguzzi},
  \citenamefont {Roschinski}, \citenamefont {Sandholzer}, \citenamefont
  {Viebahn},\ and\ \citenamefont
  {Esslinger}}]{walterQuantizationItsBreakdown2023}%
  \BibitemOpen
  \bibfield  {author} {\bibinfo {author} {\bibfnamefont {A.-S.}\ \bibnamefont
  {Walter}}, \bibinfo {author} {\bibfnamefont {Z.}~\bibnamefont {Zhu}},
  \bibinfo {author} {\bibfnamefont {M.}~\bibnamefont {G{\"a}chter}}, \bibinfo
  {author} {\bibfnamefont {J.}~\bibnamefont {Minguzzi}}, \bibinfo {author}
  {\bibfnamefont {S.}~\bibnamefont {Roschinski}}, \bibinfo {author}
  {\bibfnamefont {K.}~\bibnamefont {Sandholzer}}, \bibinfo {author}
  {\bibfnamefont {K.}~\bibnamefont {Viebahn}}, \ and\ \bibinfo {author}
  {\bibfnamefont {T.}~\bibnamefont {Esslinger}},\ }\href {\doibase
  10.1038/s41567-023-02145-w} {\bibfield  {journal} {\bibinfo  {journal}
  {Nature Physics}\ }\textbf {\bibinfo {volume} {19}},\ \bibinfo {pages} {1471}
  (\bibinfo {year} {2023})}\BibitemShut {NoStop}%
\bibitem [{\citenamefont {Viebahn}\ \emph {et~al.}(2024)\citenamefont
  {Viebahn}, \citenamefont {Walter}, \citenamefont {Bertok}, \citenamefont
  {Zhu}, \citenamefont {G{\"a}chter}, \citenamefont {Aligia}, \citenamefont
  {{Heidrich-Meisner}},\ and\ \citenamefont
  {Esslinger}}]{viebahnInteractionsEnableThouless2024}%
  \BibitemOpen
  \bibfield  {author} {\bibinfo {author} {\bibfnamefont {K.}~\bibnamefont
  {Viebahn}}, \bibinfo {author} {\bibfnamefont {A.-S.}\ \bibnamefont {Walter}},
  \bibinfo {author} {\bibfnamefont {E.}~\bibnamefont {Bertok}}, \bibinfo
  {author} {\bibfnamefont {Z.}~\bibnamefont {Zhu}}, \bibinfo {author}
  {\bibfnamefont {M.}~\bibnamefont {G{\"a}chter}}, \bibinfo {author}
  {\bibfnamefont {A.~A.}\ \bibnamefont {Aligia}}, \bibinfo {author}
  {\bibfnamefont {F.}~\bibnamefont {{Heidrich-Meisner}}}, \ and\ \bibinfo
  {author} {\bibfnamefont {T.}~\bibnamefont {Esslinger}},\ }\href {\doibase
  10.1103/PhysRevX.14.021049} {\bibfield  {journal} {\bibinfo  {journal}
  {Physical Review X}\ }\textbf {\bibinfo {volume} {14}},\ \bibinfo {pages}
  {021049} (\bibinfo {year} {2024})}\BibitemShut {NoStop}%
\bibitem [{\citenamefont {Zhu}\ \emph {et~al.}(2024)\citenamefont {Zhu},
  \citenamefont {G{\"a}chter}, \citenamefont {Walter}, \citenamefont
  {Viebahn},\ and\ \citenamefont {Esslinger}}]{Zhu2024}%
  \BibitemOpen
  \bibfield  {author} {\bibinfo {author} {\bibfnamefont {Z.}~\bibnamefont
  {Zhu}}, \bibinfo {author} {\bibfnamefont {M.}~\bibnamefont {G{\"a}chter}},
  \bibinfo {author} {\bibfnamefont {A.-S.}\ \bibnamefont {Walter}}, \bibinfo
  {author} {\bibfnamefont {K.}~\bibnamefont {Viebahn}}, \ and\ \bibinfo
  {author} {\bibfnamefont {T.}~\bibnamefont {Esslinger}},\ }\href
  {https://www.science.org/doi/10.1126/science.adg3848} {\bibfield  {journal}
  {\bibinfo  {journal} {Science}\ }\textbf {\bibinfo {volume} {384}},\ \bibinfo
  {pages} {317} (\bibinfo {year} {2024})}\BibitemShut {NoStop}%
\bibitem [{\citenamefont {Kiefer}\ \emph {et~al.}(2026)\citenamefont {Kiefer},
  \citenamefont {Zhu}, \citenamefont {Fischer}, \citenamefont {Jele},
  \citenamefont {G{\"a}chter}, \citenamefont {Bisson}, \citenamefont
  {Viebahn},\ and\ \citenamefont {Esslinger}}]{Kiefer2026}%
  \BibitemOpen
  \bibfield  {author} {\bibinfo {author} {\bibfnamefont {Y.}~\bibnamefont
  {Kiefer}}, \bibinfo {author} {\bibfnamefont {Z.}~\bibnamefont {Zhu}},
  \bibinfo {author} {\bibfnamefont {L.}~\bibnamefont {Fischer}}, \bibinfo
  {author} {\bibfnamefont {S.}~\bibnamefont {Jele}}, \bibinfo {author}
  {\bibfnamefont {M.}~\bibnamefont {G{\"a}chter}}, \bibinfo {author}
  {\bibfnamefont {G.}~\bibnamefont {Bisson}}, \bibinfo {author} {\bibfnamefont
  {K.}~\bibnamefont {Viebahn}}, \ and\ \bibinfo {author} {\bibfnamefont
  {T.}~\bibnamefont {Esslinger}},\ }\href
  {https://www.nature.com/articles/s41586-026-10285-1} {\bibfield  {journal}
  {\bibinfo  {journal} {Nature}\ }\textbf {\bibinfo {volume} {652}},\ \bibinfo
  {pages} {609} (\bibinfo {year} {2026})}\BibitemShut {NoStop}%
\bibitem [{\citenamefont {Xiao}\ \emph {et~al.}(2010)\citenamefont {Xiao},
  \citenamefont {Chang},\ and\ \citenamefont {Niu}}]{Xiao2010}%
  \BibitemOpen
  \bibfield  {author} {\bibinfo {author} {\bibfnamefont {D.}~\bibnamefont
  {Xiao}}, \bibinfo {author} {\bibfnamefont {M.-C.}\ \bibnamefont {Chang}}, \
  and\ \bibinfo {author} {\bibfnamefont {Q.}~\bibnamefont {Niu}},\ }\href
  {\doibase 10.1103/RevModPhys.82.1959} {\bibfield  {journal} {\bibinfo
  {journal} {Rev. Mod. Phys.}\ }\textbf {\bibinfo {volume} {82}},\ \bibinfo
  {pages} {1959} (\bibinfo {year} {2010})}\BibitemShut {NoStop}%
\bibitem [{\citenamefont {Essin}\ and\ \citenamefont
  {Gurarie}(2011)}]{Essin2011}%
  \BibitemOpen
  \bibfield  {author} {\bibinfo {author} {\bibfnamefont {A.~M.}\ \bibnamefont
  {Essin}}\ and\ \bibinfo {author} {\bibfnamefont {V.}~\bibnamefont
  {Gurarie}},\ }\href {\doibase 10.1103/PhysRevB.84.125132} {\bibfield
  {journal} {\bibinfo  {journal} {Phys. Rev. B}\ }\textbf {\bibinfo {volume}
  {84}},\ \bibinfo {pages} {125132} (\bibinfo {year} {2011})}\BibitemShut
  {NoStop}%
\bibitem [{\citenamefont {Huang}\ \emph {et~al.}(2025)\citenamefont {Huang},
  \citenamefont {Chen}, \citenamefont {Liang}, \citenamefont {Krebs},
  \citenamefont {Springhorn}, \citenamefont {Li}, \citenamefont {Tian},
  \citenamefont {Hazzard}, \citenamefont {Covey},\ and\ \citenamefont
  {Gadway}}]{Huang2025}%
  \BibitemOpen
  \bibfield  {author} {\bibinfo {author} {\bibfnamefont {C.}~\bibnamefont
  {Huang}}, \bibinfo {author} {\bibfnamefont {T.}~\bibnamefont {Chen}},
  \bibinfo {author} {\bibfnamefont {Q.}~\bibnamefont {Liang}}, \bibinfo
  {author} {\bibfnamefont {M.~A.}\ \bibnamefont {Krebs}}, \bibinfo {author}
  {\bibfnamefont {E.}~\bibnamefont {Springhorn}}, \bibinfo {author}
  {\bibfnamefont {R.}~\bibnamefont {Li}}, \bibinfo {author} {\bibfnamefont
  {M.}~\bibnamefont {Tian}}, \bibinfo {author} {\bibfnamefont {K.~R.~A.}\
  \bibnamefont {Hazzard}}, \bibinfo {author} {\bibfnamefont {J.~P.}\
  \bibnamefont {Covey}}, \ and\ \bibinfo {author} {\bibfnamefont
  {B.}~\bibnamefont {Gadway}},\ }\href {https://arxiv.org/abs/2512.12364}
  {\enquote {\bibinfo {title} {Interaction-assisted topological pumping in few-
  and many-atom rydberg arrays},}\ } (\bibinfo {year} {2025}),\ \Eprint
  {http://arxiv.org/abs/2512.12364} {arXiv:2512.12364 [cond-mat.quant-gas]}
  \BibitemShut {NoStop}%
\bibitem [{\citenamefont {Chen}\ \emph
  {et~al.}(2024{\natexlab{a}})\citenamefont {Chen}, \citenamefont {Huang},
  \citenamefont {Gadway},\ and\ \citenamefont {Covey}}]{Chen2024a}%
  \BibitemOpen
  \bibfield  {author} {\bibinfo {author} {\bibfnamefont {T.}~\bibnamefont
  {Chen}}, \bibinfo {author} {\bibfnamefont {C.}~\bibnamefont {Huang}},
  \bibinfo {author} {\bibfnamefont {B.}~\bibnamefont {Gadway}}, \ and\ \bibinfo
  {author} {\bibfnamefont {J.~P.}\ \bibnamefont {Covey}},\ }\href {\doibase
  10.1103/PhysRevLett.133.120604} {\bibfield  {journal} {\bibinfo  {journal}
  {Phys. Rev. Lett.}\ }\textbf {\bibinfo {volume} {133}},\ \bibinfo {pages}
  {120604} (\bibinfo {year} {2024}{\natexlab{a}})}\BibitemShut {NoStop}%
\bibitem [{\citenamefont {Chen}\ \emph
  {et~al.}(2024{\natexlab{b}})\citenamefont {Chen}, \citenamefont {Huang},
  \citenamefont {Velkovsky}, \citenamefont {Hazzard}, \citenamefont {Covey},\
  and\ \citenamefont {Gadway}}]{Chen2024}%
  \BibitemOpen
  \bibfield  {author} {\bibinfo {author} {\bibfnamefont {T.}~\bibnamefont
  {Chen}}, \bibinfo {author} {\bibfnamefont {C.}~\bibnamefont {Huang}},
  \bibinfo {author} {\bibfnamefont {I.}~\bibnamefont {Velkovsky}}, \bibinfo
  {author} {\bibfnamefont {K.~R.~A.}\ \bibnamefont {Hazzard}}, \bibinfo
  {author} {\bibfnamefont {J.~P.}\ \bibnamefont {Covey}}, \ and\ \bibinfo
  {author} {\bibfnamefont {B.}~\bibnamefont {Gadway}},\ }\href {\doibase
  10.1038/s41467-024-46823-6} {\bibfield  {journal} {\bibinfo  {journal}
  {Nature Communications}\ }\textbf {\bibinfo {volume} {15}},\ \bibinfo {pages}
  {2675} (\bibinfo {year} {2024}{\natexlab{b}})}\BibitemShut {NoStop}%
\bibitem [{\citenamefont {Chen}\ \emph
  {et~al.}(2025{\natexlab{a}})\citenamefont {Chen}, \citenamefont {Huang},
  \citenamefont {Velkovsky}, \citenamefont {Ozawa}, \citenamefont {Price},
  \citenamefont {Covey},\ and\ \citenamefont {Gadway}}]{Chen2025}%
  \BibitemOpen
  \bibfield  {author} {\bibinfo {author} {\bibfnamefont {T.}~\bibnamefont
  {Chen}}, \bibinfo {author} {\bibfnamefont {C.}~\bibnamefont {Huang}},
  \bibinfo {author} {\bibfnamefont {I.}~\bibnamefont {Velkovsky}}, \bibinfo
  {author} {\bibfnamefont {T.}~\bibnamefont {Ozawa}}, \bibinfo {author}
  {\bibfnamefont {H.}~\bibnamefont {Price}}, \bibinfo {author} {\bibfnamefont
  {J.~P.}\ \bibnamefont {Covey}}, \ and\ \bibinfo {author} {\bibfnamefont
  {B.}~\bibnamefont {Gadway}},\ }\href {\doibase 10.1038/s41567-024-02714-7}
  {\bibfield  {journal} {\bibinfo  {journal} {Nature Physics}\ }\textbf
  {\bibinfo {volume} {21}},\ \bibinfo {pages} {221} (\bibinfo {year}
  {2025}{\natexlab{a}})}\BibitemShut {NoStop}%
\bibitem [{\citenamefont {Chen}\ \emph
  {et~al.}(2025{\natexlab{b}})\citenamefont {Chen}, \citenamefont {Huang},
  \citenamefont {Covey},\ and\ \citenamefont {Gadway}}]{chen2025b}%
  \BibitemOpen
  \bibfield  {author} {\bibinfo {author} {\bibfnamefont {T.}~\bibnamefont
  {Chen}}, \bibinfo {author} {\bibfnamefont {C.}~\bibnamefont {Huang}},
  \bibinfo {author} {\bibfnamefont {J.~P.}\ \bibnamefont {Covey}}, \ and\
  \bibinfo {author} {\bibfnamefont {B.}~\bibnamefont {Gadway}},\ }\href
  {\doibase 10.1103/gb6x-m1sg} {\bibfield  {journal} {\bibinfo  {journal}
  {Phys. Rev. Lett.}\ }\textbf {\bibinfo {volume} {135}},\ \bibinfo {pages}
  {253402} (\bibinfo {year} {2025}{\natexlab{b}})}\BibitemShut {NoStop}%
\bibitem [{\citenamefont {Kanungo}\ \emph {et~al.}(2022)\citenamefont
  {Kanungo}, \citenamefont {Whalen}, \citenamefont {Lu}, \citenamefont {Yuan},
  \citenamefont {Dasgupta}, \citenamefont {Dunning}, \citenamefont {Hazzard},\
  and\ \citenamefont {Killian}}]{Kanungo2022}%
  \BibitemOpen
  \bibfield  {author} {\bibinfo {author} {\bibfnamefont {S.~K.}\ \bibnamefont
  {Kanungo}}, \bibinfo {author} {\bibfnamefont {J.~D.}\ \bibnamefont {Whalen}},
  \bibinfo {author} {\bibfnamefont {Y.}~\bibnamefont {Lu}}, \bibinfo {author}
  {\bibfnamefont {M.}~\bibnamefont {Yuan}}, \bibinfo {author} {\bibfnamefont
  {S.}~\bibnamefont {Dasgupta}}, \bibinfo {author} {\bibfnamefont {F.~B.}\
  \bibnamefont {Dunning}}, \bibinfo {author} {\bibfnamefont {K.~R.~A.}\
  \bibnamefont {Hazzard}}, \ and\ \bibinfo {author} {\bibfnamefont {T.~C.}\
  \bibnamefont {Killian}},\ }\href {\doibase 10.1038/s41467-022-28550-y}
  {\bibfield  {journal} {\bibinfo  {journal} {Nature Communications}\ }\textbf
  {\bibinfo {volume} {13}},\ \bibinfo {pages} {972} (\bibinfo {year}
  {2022})}\BibitemShut {NoStop}%
\bibitem [{\citenamefont {Trautmann}\ \emph {et~al.}(2024)\citenamefont
  {Trautmann}, \citenamefont {Sodemann~Villadiego},\ and\ \citenamefont
  {Deiglmayr}}]{Trautmann2024}%
  \BibitemOpen
  \bibfield  {author} {\bibinfo {author} {\bibfnamefont {M.}~\bibnamefont
  {Trautmann}}, \bibinfo {author} {\bibfnamefont {I.}~\bibnamefont
  {Sodemann~Villadiego}}, \ and\ \bibinfo {author} {\bibfnamefont
  {J.}~\bibnamefont {Deiglmayr}},\ }\href {\doibase
  10.1103/PhysRevA.110.L040601} {\bibfield  {journal} {\bibinfo  {journal}
  {Phys. Rev. A}\ }\textbf {\bibinfo {volume} {110}},\ \bibinfo {pages}
  {L040601} (\bibinfo {year} {2024})}\BibitemShut {NoStop}%
\bibitem [{Sup()}]{SuppMats}%
  \BibitemOpen
  \href@noop {} {}\bibinfo {note} {See Supplementary Material for more
  experimental details and details on the theoretical formulation.}\BibitemShut
  {Stop}%
\bibitem [{\citenamefont {Rubbmark}\ \emph {et~al.}(1981)\citenamefont
  {Rubbmark}, \citenamefont {Kash}, \citenamefont {Littman},\ and\
  \citenamefont {Kleppner}}]{Rubbmark1981}%
  \BibitemOpen
  \bibfield  {author} {\bibinfo {author} {\bibfnamefont {J.~R.}\ \bibnamefont
  {Rubbmark}}, \bibinfo {author} {\bibfnamefont {M.~M.}\ \bibnamefont {Kash}},
  \bibinfo {author} {\bibfnamefont {M.~G.}\ \bibnamefont {Littman}}, \ and\
  \bibinfo {author} {\bibfnamefont {D.}~\bibnamefont {Kleppner}},\ }\href
  {\doibase 10.1103/PhysRevA.23.3107} {\bibfield  {journal} {\bibinfo
  {journal} {Phys. Rev. A}\ }\textbf {\bibinfo {volume} {23}},\ \bibinfo
  {pages} {3107} (\bibinfo {year} {1981})}\BibitemShut {NoStop}%
\bibitem [{\citenamefont {Zhu}\ and\ \citenamefont {Nakamura}(1994)}]{Zhu1994}%
  \BibitemOpen
  \bibfield  {author} {\bibinfo {author} {\bibfnamefont {C.}~\bibnamefont
  {Zhu}}\ and\ \bibinfo {author} {\bibfnamefont {H.}~\bibnamefont {Nakamura}},\
  }\href@noop {} {\bibfield  {journal} {\bibinfo  {journal} {The Journal of
  chemical physics}\ }\textbf {\bibinfo {volume} {101}},\ \bibinfo {pages}
  {10630} (\bibinfo {year} {1994})}\BibitemShut {NoStop}%
\bibitem [{\citenamefont {Shih}\ and\ \citenamefont {Niu}(1994)}]{Shih1994}%
  \BibitemOpen
  \bibfield  {author} {\bibinfo {author} {\bibfnamefont {W.-K.}\ \bibnamefont
  {Shih}}\ and\ \bibinfo {author} {\bibfnamefont {Q.}~\bibnamefont {Niu}},\
  }\href {\doibase 10.1103/PhysRevB.50.11902} {\bibfield  {journal} {\bibinfo
  {journal} {Phys. Rev. B}\ }\textbf {\bibinfo {volume} {50}},\ \bibinfo
  {pages} {11902} (\bibinfo {year} {1994})}\BibitemShut {NoStop}%
\bibitem [{\citenamefont {Lindner}\ \emph {et~al.}(2017)\citenamefont
  {Lindner}, \citenamefont {Berg},\ and\ \citenamefont {Rudner}}]{Lindner2017}%
  \BibitemOpen
  \bibfield  {author} {\bibinfo {author} {\bibfnamefont {N.~H.}\ \bibnamefont
  {Lindner}}, \bibinfo {author} {\bibfnamefont {E.}~\bibnamefont {Berg}}, \
  and\ \bibinfo {author} {\bibfnamefont {M.~S.}\ \bibnamefont {Rudner}},\
  }\href {\doibase 10.1103/PhysRevX.7.011018} {\bibfield  {journal} {\bibinfo
  {journal} {Phys. Rev. X}\ }\textbf {\bibinfo {volume} {7}},\ \bibinfo {pages}
  {011018} (\bibinfo {year} {2017})}\BibitemShut {NoStop}%
\bibitem [{\citenamefont {Privitera}\ \emph {et~al.}(2018)\citenamefont
  {Privitera}, \citenamefont {Russomanno}, \citenamefont {Citro},\ and\
  \citenamefont {Santoro}}]{Privitera2018}%
  \BibitemOpen
  \bibfield  {author} {\bibinfo {author} {\bibfnamefont {L.}~\bibnamefont
  {Privitera}}, \bibinfo {author} {\bibfnamefont {A.}~\bibnamefont
  {Russomanno}}, \bibinfo {author} {\bibfnamefont {R.}~\bibnamefont {Citro}}, \
  and\ \bibinfo {author} {\bibfnamefont {G.~E.}\ \bibnamefont {Santoro}},\
  }\href {\doibase 10.1103/PhysRevLett.120.106601} {\bibfield  {journal}
  {\bibinfo  {journal} {Phys. Rev. Lett.}\ }\textbf {\bibinfo {volume} {120}},\
  \bibinfo {pages} {106601} (\bibinfo {year} {2018})}\BibitemShut {NoStop}%
\bibitem [{\citenamefont {Lunt}\ \emph {et~al.}(2024)\citenamefont {Lunt},
  \citenamefont {Hill}, \citenamefont {Reiter}, \citenamefont {Preiss},
  \citenamefont {Ga\l{}ka},\ and\ \citenamefont {Jochim}}]{Jochim-FQH}%
  \BibitemOpen
  \bibfield  {author} {\bibinfo {author} {\bibfnamefont {P.}~\bibnamefont
  {Lunt}}, \bibinfo {author} {\bibfnamefont {P.}~\bibnamefont {Hill}}, \bibinfo
  {author} {\bibfnamefont {J.}~\bibnamefont {Reiter}}, \bibinfo {author}
  {\bibfnamefont {P.~M.}\ \bibnamefont {Preiss}}, \bibinfo {author}
  {\bibfnamefont {M.}~\bibnamefont {Ga\l{}ka}}, \ and\ \bibinfo {author}
  {\bibfnamefont {S.}~\bibnamefont {Jochim}},\ }\href {\doibase
  10.1103/PhysRevLett.133.253401} {\bibfield  {journal} {\bibinfo  {journal}
  {Phys. Rev. Lett.}\ }\textbf {\bibinfo {volume} {133}},\ \bibinfo {pages}
  {253401} (\bibinfo {year} {2024})}\BibitemShut {NoStop}%
\bibitem [{\citenamefont {Wang}\ and\ \citenamefont
  {Hazzard}(2025)}]{Wang2025}%
  \BibitemOpen
  \bibfield  {author} {\bibinfo {author} {\bibfnamefont {Z.}~\bibnamefont
  {Wang}}\ and\ \bibinfo {author} {\bibfnamefont {K.~R.~A.}\ \bibnamefont
  {Hazzard}},\ }\href {\doibase 10.1038/s41586-024-08262-7} {\bibfield
  {journal} {\bibinfo  {journal} {Nature}\ }\textbf {\bibinfo {volume} {637}},\
  \bibinfo {pages} {314} (\bibinfo {year} {2025})}\BibitemShut {NoStop}%
\bibitem [{\citenamefont {Keilmann}\ \emph {et~al.}(2011)\citenamefont
  {Keilmann}, \citenamefont {Lanzmich}, \citenamefont {McCulloch},\ and\
  \citenamefont {Roncaglia}}]{Keilmann2011}%
  \BibitemOpen
  \bibfield  {author} {\bibinfo {author} {\bibfnamefont {T.}~\bibnamefont
  {Keilmann}}, \bibinfo {author} {\bibfnamefont {S.}~\bibnamefont {Lanzmich}},
  \bibinfo {author} {\bibfnamefont {I.}~\bibnamefont {McCulloch}}, \ and\
  \bibinfo {author} {\bibfnamefont {M.}~\bibnamefont {Roncaglia}},\ }\href
  {\doibase 10.1038/ncomms1353} {\bibfield  {journal} {\bibinfo  {journal}
  {Nature Communications}\ }\textbf {\bibinfo {volume} {2}},\ \bibinfo {pages}
  {361} (\bibinfo {year} {2011})}\BibitemShut {NoStop}%
\bibitem [{\citenamefont {Greschner}\ and\ \citenamefont
  {Santos}(2015)}]{Gresch-1}%
  \BibitemOpen
  \bibfield  {author} {\bibinfo {author} {\bibfnamefont {S.}~\bibnamefont
  {Greschner}}\ and\ \bibinfo {author} {\bibfnamefont {L.}~\bibnamefont
  {Santos}},\ }\href {\doibase 10.1103/PhysRevLett.115.053002} {\bibfield
  {journal} {\bibinfo  {journal} {Phys. Rev. Lett.}\ }\textbf {\bibinfo
  {volume} {115}},\ \bibinfo {pages} {053002} (\bibinfo {year}
  {2015})}\BibitemShut {NoStop}%
\bibitem [{\citenamefont {Kwan}\ \emph {et~al.}(2024)\citenamefont {Kwan},
  \citenamefont {Segura}, \citenamefont {Li}, \citenamefont {Kim},
  \citenamefont {Gorshkov}, \citenamefont {Eckardt}, \citenamefont
  {Bakkali-Hassani},\ and\ \citenamefont {Greiner}}]{joyce}%
  \BibitemOpen
  \bibfield  {author} {\bibinfo {author} {\bibfnamefont {J.}~\bibnamefont
  {Kwan}}, \bibinfo {author} {\bibfnamefont {P.}~\bibnamefont {Segura}},
  \bibinfo {author} {\bibfnamefont {Y.}~\bibnamefont {Li}}, \bibinfo {author}
  {\bibfnamefont {S.}~\bibnamefont {Kim}}, \bibinfo {author} {\bibfnamefont
  {A.~V.}\ \bibnamefont {Gorshkov}}, \bibinfo {author} {\bibfnamefont
  {A.}~\bibnamefont {Eckardt}}, \bibinfo {author} {\bibfnamefont
  {B.}~\bibnamefont {Bakkali-Hassani}}, \ and\ \bibinfo {author} {\bibfnamefont
  {M.}~\bibnamefont {Greiner}},\ }\href {\doibase 10.1126/science.adi3252}
  {\bibfield  {journal} {\bibinfo  {journal} {Science}\ }\textbf {\bibinfo
  {volume} {386}},\ \bibinfo {pages} {1055} (\bibinfo {year}
  {2024})}\BibitemShut {NoStop}%
\end{thebibliography}%

\clearpage

\renewcommand{\thesection}{\Alph{section}}
\renewcommand{\thefigure}{S\arabic{figure}}
\renewcommand{\thetable}{S\Roman{table}}
\setcounter{figure}{0}
\renewcommand{\theequation}{S\arabic{equation}}
\renewcommand{\thepage}{S\arabic{page}}
\setcounter{equation}{0}
\setcounter{page}{1}

\begin{widetext}
\appendix

\section{\large Supplemental Material for ``Interaction-enabled topological \\ pumping of Rydberg electrons"}

\vspace{5mm}

\subsection{Mapping two-color-microwave driven Rydberg-state chain into 
Rice-Mele model in atom-pair basis}

We consider two interacting multi-level Rydberg atoms (labeled $A$ and $B$), with each pair of adjacent Rydberg states coupled by a slowly modulated bichromatic microwave field. As shown in Fig.~1 in the main text, for $\ket{n}\leftrightarrow\ket{n+1}$ with the transition frequency $\omega_n$, the time-dependent hopping rates for the blue- and red-detuned tones are given by
\begin{equation}\label{eqS1}
 J_{\pm} = \frac{{J}_0}{2\sqrt{2}} \left[1\pm\sin{(\omega t)}\right]^2,
\end{equation}
respectively. The corresponding phase modulations are
\begin{equation}\label{eqS2}
 \theta_{\pm} = \pm\Delta_c t \pm \frac{\Delta}{\omega}\sin{(\omega t)},
\end{equation}
where $\Delta_c$ denotes a constant frequency shift to the two tones, and $\Delta$ is the time-dependent detuning amplitude. The full Hamiltonian reads
\begin{equation}\label{eqS3}
 H = H_0 + H_{\rm mw} + H_{\rm dip},
\end{equation}
in which
\begin{eqnarray}
 H_0 &=& \sum_n \sum_{\alpha\in\{A,B\}}\epsilon_n \ket{n}_\alpha\bra{n}, \nonumber \\
 H_{\rm mw} &=& \sum_n \sum_{\alpha\in\{A,B\}}\left[J_+ e^{i(\omega_n t+\theta_+)} + J_- e^{i(\omega_n t +\theta_-)} + {\rm c.c.}\right]\ket{n}_\alpha\bra{n+1} + {\rm H.c.} \\
 H_{\rm dip} &=& \sum_n V_n \ket{n+1}_A\bra{n} \otimes \ket{n}_B\bra{n+1} + {\rm H.c.}, \nonumber
\end{eqnarray}
with $\omega_n = \epsilon_{n+1}-\epsilon_{n}$ and $V_n$ the dipolar exchange strength for Rydberg states $\ket{n}$ and $\ket{n+1}$. 

Due to the relatively slow time variations of $J_\pm$ and $\theta_\pm$, i.e., $\omega \ll \omega_n$, we arrive at the Hamiltonian under the interaction picture by applying a unitary transformation as
\begin{eqnarray}
 H_{\rm int} &=& e^{iH_0 t}(H_{\rm mw} + H_{\rm dip}) e^{-iH_0 t} \nonumber \\
  & = & \sum_n \sum_{\alpha\in\{A,B\}}(J_+ e^{i\theta_+} + J_- e^{i\theta_-})\ket{n}_\alpha\bra{n+1} + \sum_n V_n \ket{n+1}_A\bra{n} \otimes \ket{n}_B\bra{n+1} + {\rm H.c.}.
\end{eqnarray}

Then, we move to truncated pair basis by only considering the triplet states for each adjacent Rydberg state pair, and label
\begin{equation}
 \ket{a_n} = \ket{n}_A\ket{n}_B, ~ \ket{b_n} = \ket{+}_{n}= \frac{\ket{n}_A\ket{n+1}_B + \ket{n+1}_A\ket{n}_B}{\sqrt{2}}.
\end{equation}
The Hamiltonian can be transformed into 
\begin{equation}
 H_{\rm pair} = \sum_n\left[(J_1 e^{i\theta_+} + J_2 e^{i\theta_-})(\ket{a_n}\bra{b_n}+\ket{b_n}\bra{a_{n+1}}) + {\rm H.c.}\right] + \sum_n V_n\ket{b_n}\bra{b_n},
\end{equation}
where the collective hopping rates $J_1=\sqrt{2}J_+$ and $J_2=\sqrt{2}J_-$. 

Next, we perform a local time-dependent unitary gauge transformation
\begin{equation}
 U = \exp\left(-i\sum_n (\xi_n\ket{a_n}\bra{a_n} + \chi_n\ket{b_n}\bra{b_n})\right),
\end{equation}
with the phases $\xi_n = \theta_+$ and $\chi_n = 0$. By noting that $\theta_+ +\theta_- = 0$, we arrive at the effective Hamiltonian
\begin{eqnarray}
 H_{\rm eff} &=& U H_{\rm pair} U^\dagger + i\partial_t U U^\dagger \nonumber\\
 &=& \sum_n\left[(J_1 + J_2e^{i 2\theta_-})\ket{a_n}\bra{b_n}+(J_1e^{i 2\theta_+}+J_2)\ket{b_n}\bra{a_{n+1}}+{\rm H.c.}\right] + \sum_n V_n\ket{b_n}\bra{b_n} + \frac{\partial \theta_+}{\partial t}\sum_n\ket{a_n}\bra{a_n}.
\end{eqnarray}
By noting that $\Delta_c, V_n \gg \Delta, J_1, J_2, \omega$ in our experiment, we neglect the fast oscillating $e^{i 2\theta_\pm}\approx e^{\pm i 2\Delta_c t}$ terms and obtain
\begin{equation}
 H_{\rm eff} = \sum_n\left(J_1\ket{a_n}\bra{b_n} + J_2\ket{b_n}\bra{a_{n+1}}+{\rm H.c.}\right) + \sum_n V_n \ket{b_n}\bra{b_n} + \sum_n [\Delta_c + \Delta \cos{(\omega t)}] \ket{a_n}\bra{a_n},
\end{equation}
where $J_1 = \frac{J_0}{2} \left[1 + \sin{(\omega t)}\right]^2$ and $J_2 = \frac{J_0}{2} \left[1 - \sin{(\omega t)}\right]^2$. Defining $a_n^\dagger$ ($b_n^\dagger$) as the creation operator for the site $\ket{a_n}$ ($\ket{b_n}$), we finally arrive at the effective Rice-Mele Hamiltonian (2) in the main text.

While the effective Rice-Mele Hamiltonian stands under truncated pair basis, we note that additional leakage channels to other long-range pair states are present in our actual experimental setting, e.g., transition from the triplet $\ket{+}_n=(\ket{n}_A\ket{n+1}_B + \ket{n+1}_A\ket{n}_B)/\sqrt{2}$ to the state $(\ket{n}_A\ket{n+2}_B + \ket{n+2}_A\ket{n}_B)/\sqrt{2}$. This is evidenced by the simulation results with the original full Hamiltonian (\ref{eqS3}) in Fig.~3(a) in the main text. Over one modulation period, the simulation curve for population in $\ket{2}$ state slightly increases to $\sim 0.1$, indicating the leakage out of the truncated Hilbert space from $\ket{+}_0$ to $(\ket{0}_A\ket{2}_B + \ket{2}_A\ket{0}_B)/\sqrt{2}$. Although the desired transition $\ket{+}_0 \to \ket{1,1}$ still dominates the second half pumping cycle, such leakage processes definitely mess up the pumping dynamics and degrade the observed pumping efficiency.

\subsection{Determining the critical modulation frequency via Landau-Zener adiabatic condition}

The pair pumping within one period can be treated as two adiabatic transfer processes: $\ket{a_n}\to\ket{b_n}$ in the first half modulation cycle, and $\ket{b_n}\to\ket{a_{n+1}}$ in the second half cycle. As the pumping trajectories for the two half cycles are symmetric in our experiment [See Figs.~1(c,d) in the main text], here we only consider the adiabatic condition for the first half cycle. By encoding $\ket{\downarrow}=\ket{a_n}$ and $\ket{\uparrow}=\ket{b_n}$, the pumping dynamics are effectively governed by 
\begin{equation}
 H = \frac{\Omega(t)}{2}\sigma_x + \frac{\delta(t)}{2}\sigma_z.
\end{equation}
The general adiabatic condition for this two-level Landau-Zener-type state transfer is
\begin{equation}
 \left|\frac{\partial \theta(t)}{\partial t}\right| \ll g(t),
\end{equation} 
with $\tan{\theta (t)} = \Omega (t)/\delta(t)$ and the instantaneous eigenenergy gap $g(t)  = \sqrt{\delta(t)^2 + \Omega(t)^2}$. Here we use the ratio 
\begin{equation}
 \xi = \left|\frac{\partial \theta(t)}{\partial t}\right| / g(t) = \frac{|\delta \frac{\partial\Omega}{\partial t} - \Omega \frac{\partial\delta}{\partial t}|}{(\delta^2 + \Omega^2)^{3/2}} = 1
\end{equation}
to determine the upper and lower critical modulating frequencies $\omega_u$ and $\omega_l$. 

Considering the case $\Delta_c=V$ and using the experimental driving protocol
\begin{eqnarray}
 \Omega(t) &=& J_0 [1+\sin{(\omega t)}]^2,\nonumber\\
 \delta(t) &=& \Delta \cos{(\omega t)},
\end{eqnarray}
we have the instantaneous adiabatic parameter 
\begin{equation}
 \xi (t) = \frac{J_0\Delta\omega[1+\sin{(\omega t)}]^2[2-\sin{(\omega t)}]}{[\Delta^2\cos^2(\omega t)+J_0^2(1+\sin{(\omega t)})^4]^{3/2}}.
\end{equation}
The state transfer starts working when the minimum value of $\xi (t)$ is smaller than 1, i.e., 
\begin{equation}
 {\rm min} ~\xi (t) \ll 1, {\rm for~} t\in [0, \pi/\omega]. 
\end{equation}
The minimum value occurs at $t = 0$ and $t=\pi/\omega$, and we have 
\begin{equation}
 \xi (t=0) = \frac{2J_0\Delta \omega}{(\Delta^2 + J_0^2)^{3/2}} \ll 1
\end{equation}
This leads to the weak LZ adiabatic critical modulation frequency 
\begin{equation}
 \omega_u = \frac{(\Delta^2 + J_0^2)^{3/2}}{2J_0\Delta} \approx \frac{\Delta^2}{2J_0}
\end{equation}
for $\Delta \gg J_0$. On the other hand, the state transfer process becomes fully adiabatic when the maximum value of $\xi (t)$, which occurs at the resonance point $t=\pi/2\omega$, is much smaller than 1, i.e.,
\begin{equation}
 \xi (t = \pi/2\omega) = \frac{\Delta\omega}{16J_0^2} \ll 1.
\end{equation}
This gives the tight LZ adiabatic critical value for the modulating frequency
\begin{equation}
 \omega_l = \frac{16J_0^2}{\Delta},
\end{equation}
below which an effective quantized state transfer shall happen subsequently in order. These two values constrain the effective charge pumping in our experiments even when the parameters of our system are such that it should be operating in the non-trivial topological regime. 

\subsection{Additional experimental details: State information, parameter modulation, and data renormalization}

We use five Rydberg states to build the synthetic lattice, i.e., $\{42S_{1/2}, 42P_{3/2}, 43S_{1/2}, 43P_{3/2}, 44S_{1/2}\}$, all with $m_J=1/2$ and indexed by $\ket{n=0}-\ket{4}$. Each transition $\ket{n}\to\ket{n+1}$ between neighboring states (energy gap $\omega_n$) can be individually addressed by microwave spectrum engineering. For each frequency tone, we also introduce the amplitude and phase modulations, based on Eq.~(\ref{eqS1}) ans Eq.~(\ref{eqS2}), to achieve an effective time-dependent Rice-Mele Hamiltonian. The exposure time window of our microwave pulse is $\sim 3.3~\mu{\rm s}$ in this experiment, which results in a lower limit of the modulation frequency as $\sim 2\pi\times 0.3~{\rm MHz}$.

The state-dependent dipolar exchange interactions $V_n=-C_3^{(n,n+1)}/(2d^3)$, with $d$ the interatomic spatial separation and 
\begin{equation}
 C_3^{(n,n+1)} = h\times \{-1502, -1289, -1657, -1422\}~{\rm MHz~\mu m^3}\quad {\rm for}~ n=0, 1, 2, 3. 
\end{equation}
Since $V_n$ shares the same scaling as $d$, we simply define $V\equiv V_0 = -C_3^{01}/(2d^3)$ to characterize the interaction strength in the effective model. We note that these non-uniform $C_3$ values lead to $\sim$15\% variation of the interactions for different state pairs, which slightly affects the pumping dynamics and efficiency when the global non-trivial topological feature maintains for all $V_n$ in our experiment. 

Following the analysis of Ref. \cite{Chen2024}, the experimentally measured populations exhibit a finite contrast lower than that of the renormalized data used in the main text. Specifically, the measurements are characterized by an average upper ceiling ($P_u$) and a lower background baseline ($P_l$). The upper ceiling mainly originates from atom loss during the imaging sequence and imperfect depletion of Rydberg-state populations. The finite background baseline is attributed to two-photon STIRAP transfer inefficiency as well as spontaneous decay and subsequent recapture of short-lived Rydberg states. These processes allow a fraction of atoms remaining in Rydberg states to contribute to the fluorescence signal, thereby reducing the overall measurement contrast. The measured average population $P_{n}^{\rm bare}$ in state $\ket{n}$ is converted to the renormalized value $P_n = (P_n^{\rm bare}-P_l)/(P_u-P_l)$, with the experimentally determined $P_u = 0.93(1)$ and $P_l=0.33(3)$ in this work. Since $P_l$ depends on the decay of Rydberg states, all experiments are performed with a fixed evolution time window of $\sim 5~\mu{\rm s}$ between the initial Rydberg excitation and subsequent de-excitation. As a result, the detection sequence, based on fluorescence imaging of ground-state atoms, always occurs at the same point in time, ensuring a consistent baseline offset for all measurements.

\clearpage
\end{widetext}
\end{document}